\documentclass[]{aastex631}

\usepackage{array}
\usepackage{amsmath}
\usepackage{booktabs}
\usepackage{multirow}

\usepackage[normalem]{ulem} % use normalem to protect \emph
\newcommand\hllightgray{\bgroup\markoverwith
  {\textcolor{lightgray}{\rule[-.5ex]{2pt}{2.5ex}}}\ULon}

\begin{document}

\title{Identification of 4FGL uncertain sources at Higher Resolutions with Inverse Discrete Wavelet Transform}

\author[0000-0002-8626-8686]{Haitao Cao}
\affiliation{School of Information Engineering, Guangzhou Panyu Polytechnic, Guangzhou, 511483, China}

\author[0000-0001-8244-1229]{Hubing Xiao}
\affiliation{Shanghai Key Lab for Astrophysics, Shanghai Normal University, Shanghai, 200234, China}

\author{Zhijian Luo}
\affiliation{Shanghai Key Lab for Astrophysics, Shanghai Normal University, Shanghai, 200234, China}

\author{Xiangtao Zeng}
\affiliation{Center for Astrophysics, Guangzhou University, Guangzhou 510006, China}
\affiliation{Astronomy Science and Technology Research Laboratory of Department of Education of Guangdong Province, Guangzhou 510006, China}
\affiliation{Key Laboratory for Astronomical Observation and Technology of Guangzhou, Guangzhou 510006, China}

\author[0000-0002-5929-0968]{Junhui Fan}
\affiliation{Center for Astrophysics, Guangzhou University, Guangzhou 510006, China}
\affiliation{Astronomy Science and Technology Research Laboratory of Department of Education of Guangdong Province, Guangzhou 510006, China}
\affiliation{Key Laboratory for Astronomical Observation and Technology of Guangzhou, Guangzhou 510006, China}

\correspondingauthor{Hubing Xiao}
\email{hubing.xiao@shnu.edu.cn}

\begin{abstract}
In the forthcoming era of big astronomical data, it is a burden to find out target sources from ground-based and space-based telescopes.
{Although Machine Learning (ML) methods have been extensively utilized to address this issue, the incorporation of in-depth data analysis can significantly enhance the efficiency of identifying target sources when dealing with massive volumes of astronomical data.}
In this work, we focused on the task of finding AGN candidates and identifying BL Lac/FSRQ candidates from the 4FGL\_DR3 uncertain sources.
We studied the correlations among the attributes of the 4FGL\_DR3 catalogue and proposed a novel method, named FDIDWT, to transform the original data.
The transformed dataset is characterized as low-dimensional and feature-highlighted, with the estimation of correlation features by Fractal Dimension (FD) theory and the multi-resolution analysis by Inverse Discrete Wavelet Transform (IDWT).
Combining the FDIDWT method with an improved lightweight MatchboxConv1D model, we accomplished two missions: 
(1) to distinguish the Active Galactic Nuclei (AGNs) from others (Non-AGNs) in the 4FGL\_DR3 uncertain sources with an accuracy of {$96.65\% \pm 1.32\%$}, namely, Mission A; 
(2) to classify blazar candidates of uncertain type (BCUs) into BL Lacertae objects (BL Lacs) or Flat Spectrum Radio Quasars (FSRQs) with an accuracy of {$92.03\% \pm 2.2\%$}, namely, Mission B. There are {1354} AGN candidates in Mission A, {482} BL Lacs candidates and {128} FSRQ candidates in Mission B were found.
The results show a high consistency of greater than {98\%} with the results in previous works.
In addition, our method has the advantage of finding less variable and relatively faint sources than ordinary methods.

\end{abstract}


\section{Introduction}
\label{sect_introduction}
Active Galactic Nucleus (AGN) has been a hot topic for more than sixty years in astronomy since its discovery in 1963 \citep{Schmidt1963Nature197}.
It is believed that AGNs centred a supermassive black hole (SMBH), which is surrounded by an accretion disk, the whole system provides energy for AGN radiation \citep{Lynden-Bell1969, Blandford1977MNRAS, Blandford1982}.
The emissions from AGNs are observed to span over the entire electromagnetic spectrum, these emissions are found to be strong and variable.
Based on the ratio of radio emission strength to the optical one, AGNs are divided into radio-loud and radio-quiet ones \citep{Strittmatter1980, Kellermann1989}, and this method has recently been modified by a `double-criterion' method \citep{Xiao2022PASJ}.
Blazars, a subclass of radio-loud AGNs that with jets pointing toward the observer, show multi-bands high and fast variability, high and variable polarization, strong and variable $\gamma$-ray emissions, and apparent superluminal motion \citep{Wills1992, Urry1995, Villata2006, Fan2002, Fan2014, Fan2021, Gupta2016, Xiao2019, Xiao2020, Xiao2022MNRAS, Abdollahi2020}.
{Blazars consist of BL Lacertae object (BL Lac)} and flat spectrum radio quasar (FSRQ), the former one shows strong emission lines (rest-frame equivalent width, ${\rm EW > 5 \AA}$) and the latter one demonstrates no or weak emission features (${\rm EW < 5 \AA}$) \citep{Urry1995, Scarpa1997}.

The study of blazars was severely limited due to the small sample size, until the launch of the Large Area Telescope on board the \textit{Fermi} Gamma-Ray Space Observatory (\textit{Fermi}-LAT) in 2008. 
It has an unprecedented performance, a better energy resolution, angular resolution, and a wider effective area in both low-energy and high-energy bands\footnote{{\url{https://www.slac.stanford.edu/exp/glast/groups/canda/lat\_Performance.htm}}} than its predecessor \textit{EGRET} \citep{Thompson1993ApJS86}, at the energy ranges from 20 MeV to 300 GeV.
\textit{Fermi}-LAT collaboration has released 5 main $\gamma$-ray sources catalogues, namely 0FGL, 1FGL, 2FGL, 3FGL and 4FGL \citep{Abdo2009ApJS183, Abdo2010ApJS188, Nolan2012, Acero2015, Abdollahi2020}.
However, there are still significant \textit{Fermi} sources not related to any known class, e.g., 1010 unassociated sources in 3FGL and 2291 uncertain sources (2157 unassociated sources + 134 unknown sources) in the latest 4FGL\_DR3, 573 blazar candidate of uncertain types (BCUs) in the 3FGL and 1493 BCUs in the latest 4FGL\_DR3.

It is time-consuming to verify these uncertain sources one by one through optical observation, thus more efficient methods must be explored. 
Many machine-learning-based algorithms have been employed for this issue. 
For instance, \cite{SazParkinson2016} applied two different Machine Learning (ML) methods (Random Forest (RF) and Logistic Regression (LR)) to identify 1008 unassociated sources in 3FGL. 
Among them, 334 sources were predicted as Pulsars (PSRs), and 559 sources were predicted as AGNs. 
\cite{Chiaro2016} utilized Blazar Flaring Patterns (B-FlaP) as an identification approach on BCUs. 
Since variability is one of the characterizing properties of blazar \citep{Paggi2011ApJ736}, the light curve of blazar was used by an Artificial Neural Networks (ANN) to identify 573 BCUs in 3FGL. 
These BCUs were associated with 342 BL Lacs and 154 FSRQs, and 77 sources remained uncertain.
\cite{Xiao2020AC} carried out an ensemble ML method, picked out 748 AGNs candidates from 1010 3FGL unassociated sources, and identified 573 BCUs to be 326 BL Lac candidates and 247 FSRQ candidates.
Moreover, \cite{Kang2019ApJ887} studied the classification of 1312 BCUs from 4FGL\_DR1 via three supervised ML methods and obtained 724 BL Lac and 332 FSRQ candidates.

The task of \textit{Fermi} source classification can be seen as feature extraction with ML methods due to their ability to learn patterns from data and provide valuable insights, decisions, and predictions \citep{jordan2015machine, zhou2017machine}. 
However, traditional ML methods are limited when dealing with the increasing volume of big astronomical data with the successful launch of more and more telescopes and detectors. 
In recent years, the popularity of Graphics Processing Units (GPUs) fires the research of Deep Learning (DL) for learning features from massive-scale data using Deep Neural Networks (DNNs). It has become a major research focus in the ML field \citep{yu2010deep, liu2017survey}. 
DNNs have proven to be successful in various real-world applications \citep{jifara2019medical, zyner2019naturalistic, alemany2019predicting, lam2019gaussian, chen2018iterative}. 
Furthermore, it has been shown that more complex problems require deeper networks \citep{bengio2009learning,he2016deep}, which has led to the development of sophisticated networks such as VGG \citep{Simonyan2014VeryDC}, ResNet \citep{he2016deep}, and ChatGPT \citep{chatgpt}.

However, rather than solely focusing on the deep structure of networks that helps to learn \textit{intrinsic features}, attribute analysis should be emphasized as well. We propose that {there exists a \textit{correlation feature}} among the attributes of raw data, and it will improve the learning performance further. In addition, numerous studies have shown that real-world data often contain highly redundant and unimportant attributes \citep{bakshi1993wave, bengio2009learning, glorot2011deep}. This redundancy can lead to sparsity in high-dimensional space, where most samples in the dataset are far away from each other. In classification tasks, this sparsity can result in less reliable predictions than in low dimensions, as predictions are based on larger extrapolations \citep{geron2017hands}. Therefore, we believe that attribute analysis presents an opportunity for better results through additional correlation features and dimension reduction.

In this paper, we focus on two missions for 4FGL\_DR3, i.e., classifying 2291 uncertain sources into AGN or Non-AGN and associating 1493 BCUs to BL Lac or FSRQ, which are named Mission A and Mission B, respectively. 
Firstly, we study some popular attribute analysis methods and glimpse the attributes of 4FGL\_DR3 sources. Then we find out the core attributes based on the Fractal Dimension (FD) theory and step into the research of multi-attribute analysis from the perspective of the whole dataset. Based on the results, we propose a novel method called FDIDWT, which is the combination of FD and Inverse Discrete Wavelet Transform (IDWT), to extract correlation features at a higher resolution. By FDIDWT, the original dataset is transformed into a low-dimensional and feature-highlighted set, which benefits the subsequent learning process. In the end, we combine the FDIDWT method with a lightweight Convolutional Neural Network (CNN) model as our contribution to accomplish the classification missions.

The paper is organized as follows. Sec. \ref{sect_data_prepro} describes the datasets in two missions and presents some {commonly used} attribute analysis methods. Based on that, Sec. \ref{sect_proposed} interprets our proposed method in detail. The experiments and results are reported in Sec. \ref{sect_exps}. Further discussions and conclusions are presented in Sec. \ref{sect_discussion} and \ref{sect_conclusion}, respectively.

\section{Datasets and Attribute Analysis}
\label{sect_data_prepro}

\subsection{Samples of 4FGL\_DR3}
\label{subsect_data_4fgl}
The \textit{Fermi}-LAT collaboration has recently released the incremental version of the 12-year \textit{Fermi}-LAT Gamma-ray Source Catalog (4FGL\_DR3, \citealp{Abdollahi2022ApJS260}). 
It contains 6659 sources, among which 3809 sources are AGNs, 559 sources are associated with Non-AGNs (including pulsars, high-mass binaries, and supernova remnants, etc.), and 2291 uncertain sources (134 sources are associated with counterparts of unknown nature and 2157 unassociated sources).
Within the AGNs, there are 3743 sources confirmed with blazars, among which 1456 blazars are associated with BL Lacs, 794 blazars are associated with FSRQs, and the rest of 1493 Blazars are BCUs that have not been tagged as BL Lacs or FSRQ. 

To accomplish Mission A and B mentioned above, {we need to select features that can distinguish one from another. Variability is one of the widely known characteristics of AGNs with respect to others that are detected by \textit{Fermi}-LAT \citep{Abdollahi2020, Abdollahi2022ApJS260}, those features that contain variability information (`Flux1000', `Flux\_Band', `Variability\_Index' and `Frac\_Variability') should be included for accomplishing Mission A. While FSRQs and BL Lacs demonstrate significant $\gamma$-ray spectra, the GeV $\gamma$-ray regime situates at the different regimes of the higher hump of blazar spectral energy distribution (SED) \citep{Fan2016, Yang2022RAA22, Yang2023SCPMA}, those features that contain spectral information (`Flux\_Band', `Pivot\_Energy', `PL\_Index') should be included for Mission B.}

In this case, we compile the data of 13 attributes from 4FGL\_DR3, as listed in Tab. \ref{tab_features}.

\begin{table}[h]
\footnotesize
\centering
\caption{The 13 selected attributes from 4FGL\_DR3 \label{tab_features}}
\begin{tabular}{rll}
\specialrule{0em}{2pt}{2pt}\toprule\midrule
\textbf{Index} & \textbf{Attribute} & \textbf{Description} \\ 
\midrule
$a_1$ & Pivot\_Energy      & Energy at which error on differential flux is minimal \\
$a_2$ & Flux1000           & Integral photon flux from 1 to 100 GeV \\
$a_3$ & PL\_Index          & Best fit power-law index \\
$a_4$ & Variability\_Index & Sum of 2A-log(Likelihood) difference between the flux fitted in each \\
   &                    & time interval and the average flux over the full catalog interval \\  
$a_5$ & Frac\_Variability  & Fractional variability computed from the fluxes in each year \\
$a_6$ & Flux\_Band1        & Integral photon flux in the spectral band 0.05 - 0.1 GeV \\
$a_7$ & Flux\_Band2        & Integral photon flux in the spectral band 0.1 - 0.3 GeV \\
$a_8$ & Flux\_Band3        & Integral photon flux in the spectral band 0.3 - 1 GeV \\
$a_9$ & Flux\_Band4        & Integral photon flux in the spectral band 1 - 3 GeV \\
$a_{10}$ & Flux\_Band5       & Integral photon flux in the spectral band 3 - 10 GeV \\
$a_{11}$ & Flux\_Band6       & Integral photon flux in the spectral band 10 - 30 GeV \\
$a_{12}$ & Flux\_Band7       & Integral photon flux in the spectral band 30 - 100 GeV \\
$a_{13}$ & Flux\_Band8       & Integral photon flux in the spectral band 100 - 1000 GeV \\
\bottomrule
\end{tabular}
\end{table}

% As to Mission A and Mission B, we split the data into three splits with a ratio of 8:1:1, as shown in Fig. \ref{fig_datasets}.  The training set is used to fit the ML methods or DNN models. The validation set aids in fine-tuning hyper-parameters for better performance. The test set is independent to evaluate the generalization capability. In this section, we analyze the attributes of the training set with the methods shown in Fig. \ref{fig_attri_analysis}.

% It is emphasized that we associate the sample to different parts according to the hash of source name \citep{speechcommandsv2} and determine which set it should belong to. With this method, the samples will be kept in the changeless training, validation, or test sets, which is good for fair comparison with other work.

\begin{figure}[htbp]
\centering
\includegraphics[scale=0.4]{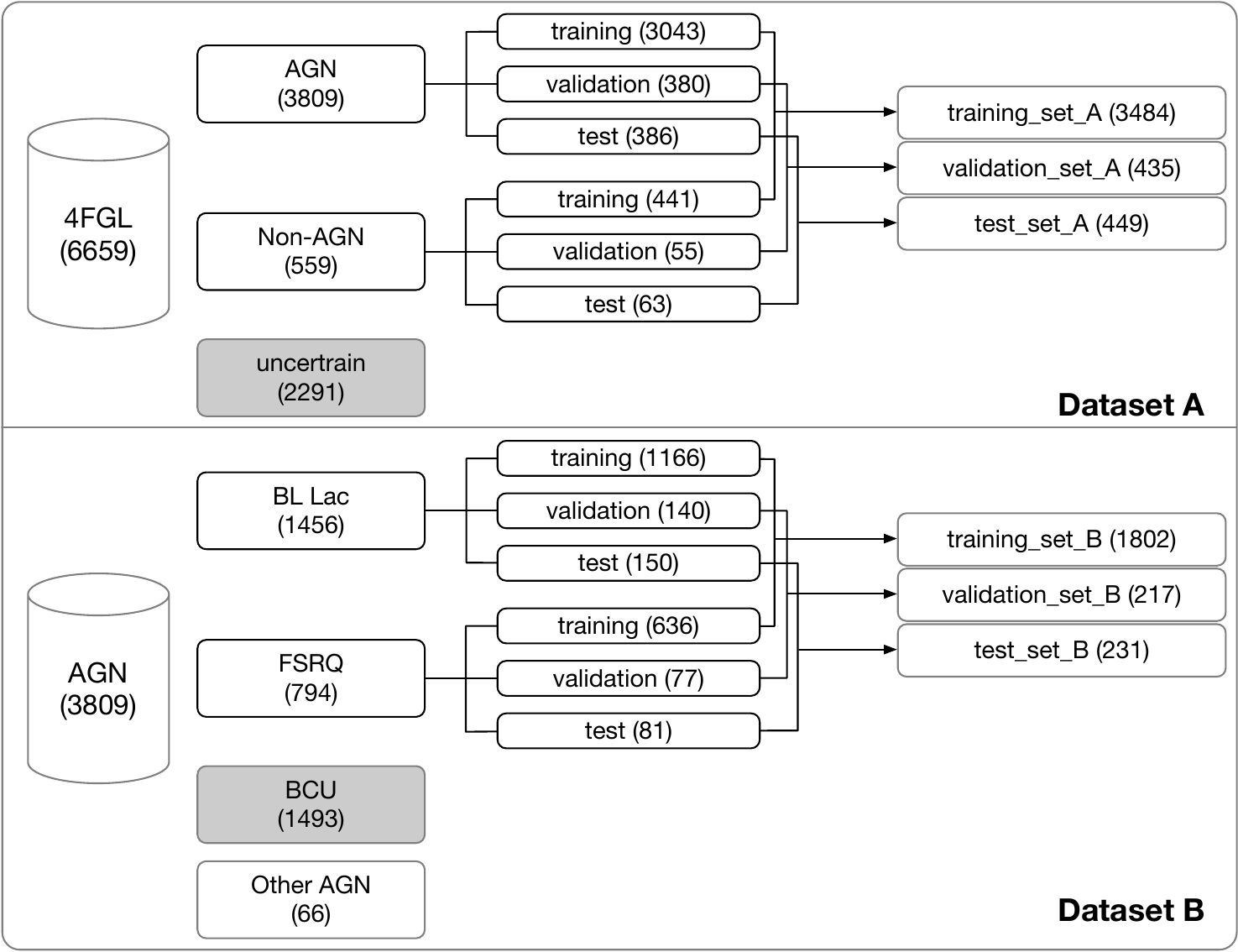}
\caption{The Dataset A (upper penale) and Dataset B (lower panel) for Mission A and Mission B, respectively.}
\label{fig_datasets}
\end{figure}

\begin{figure}[htbp]
\centering
\includegraphics[scale=0.4]{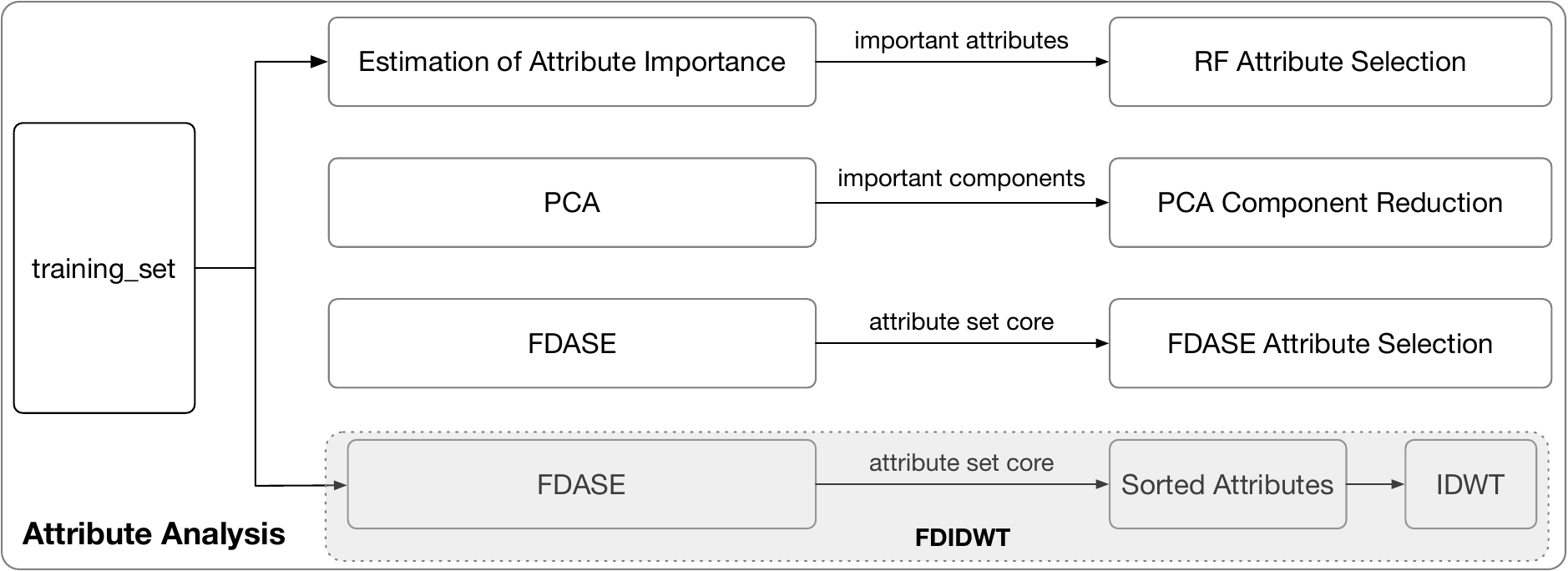}
\caption{The commonly used 3 attribute analysis methods and our proposed method (grey color).}
\label{fig_attri_analysis}
\end{figure}

{As to Mission A and Mission B, we randomly split the data into three subsets with a ratio of 8:1:1, as shown in Fig. \ref{fig_datasets}. The training set is used to fit the ML methods or DNN models, and the validation set aids in fine-tuning hyper-parameters for better performance. Finally, the generalization capability of the model is evaluated on the test set independently. Moreover, to explore the robustness of the model, this split policy was repeated 10 times and in the end, 10 datasets were prepared for each mission. In this section, we analyze the attributes of the training set with the methods shown in Fig. \ref{fig_attri_analysis}.}

\subsection{Attribute Importance}
\label{subsect_attri_impo}

As mentioned before, it is found that real-world data often contain highly redundant and unimportant attributes. The Decision Tree (DT) is a popular technique to estimate the importance of attributes. The attributes that appear in the tree are considered important, with the frequency of their appearance being their attribute importance. The less frequently an attribute appears, the less important it is assumed to be. RF is composed of multiple DTs and reduces the bias in estimating attribute importance \citep{breiman2001random}. It is always the practice to remove the unimportant attributes and perform the so-called Attribute Selection (AS) for dimension reduction.

\begin{figure}[htbp]
\centering
\includegraphics[scale=0.35]{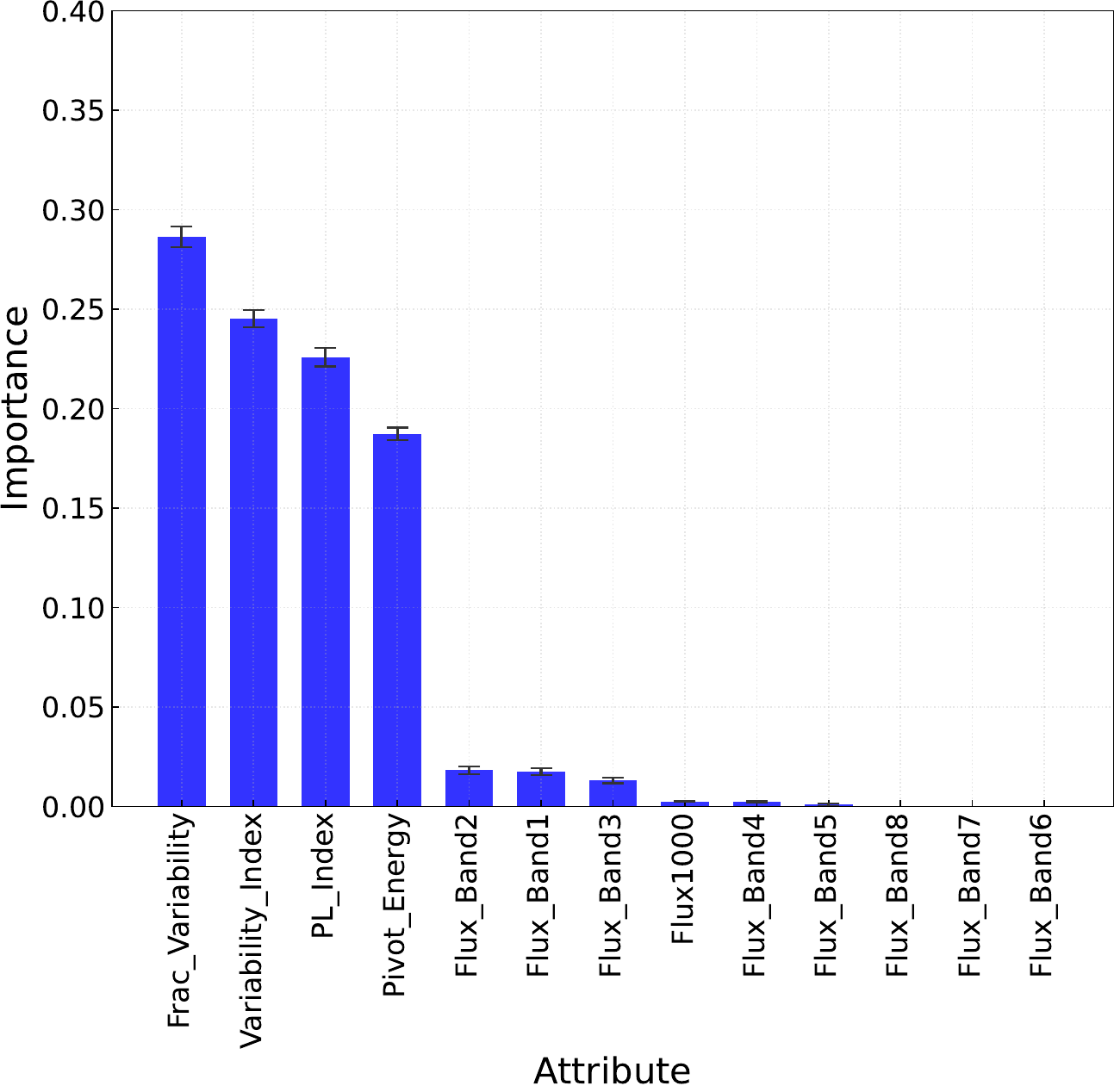}
% \hspace{0.5cm}
\includegraphics[scale=0.35]{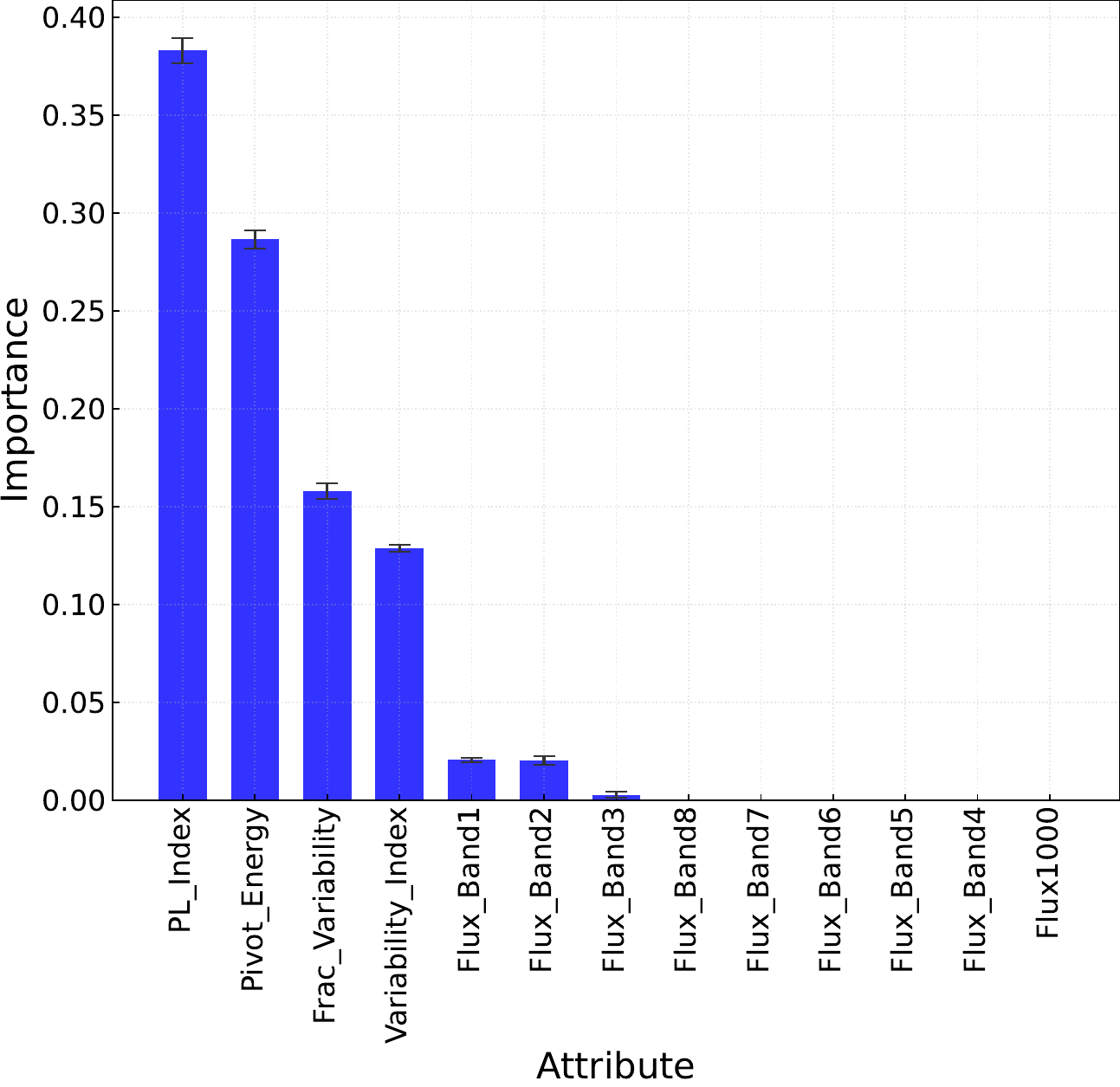}
% \hspace{0.5cm}
\caption{The {average} attribute importance of the training sets in Dataset A (left) and Dataset B (right).}
\label{fig_attri_importance}
\end{figure}

We employ 50000 DTs to build the RF and take the entropy as the criterion. Other hyper-parameters align to the default values of scikit-learn \citep{pedregosa2011scikit}. The results of attribute importance on the training sets are {averaged over 10 datasets and} shown in Fig. \ref{fig_attri_importance}. Interestingly, from the results, we find that {the four most important attributes} are `Pivot\_Energy', `PL\_Index', `Variability\_Index', and `Frac\_Variability' both in Dataset A and B.
%Thus in the following experiments, we can compare some trials on the reduced datasets with different dimensions.

\subsection{Principal Component Analysis}
\label{subsect_pca}
\begin{figure}[htbp]
\centering
\includegraphics[scale=0.35]{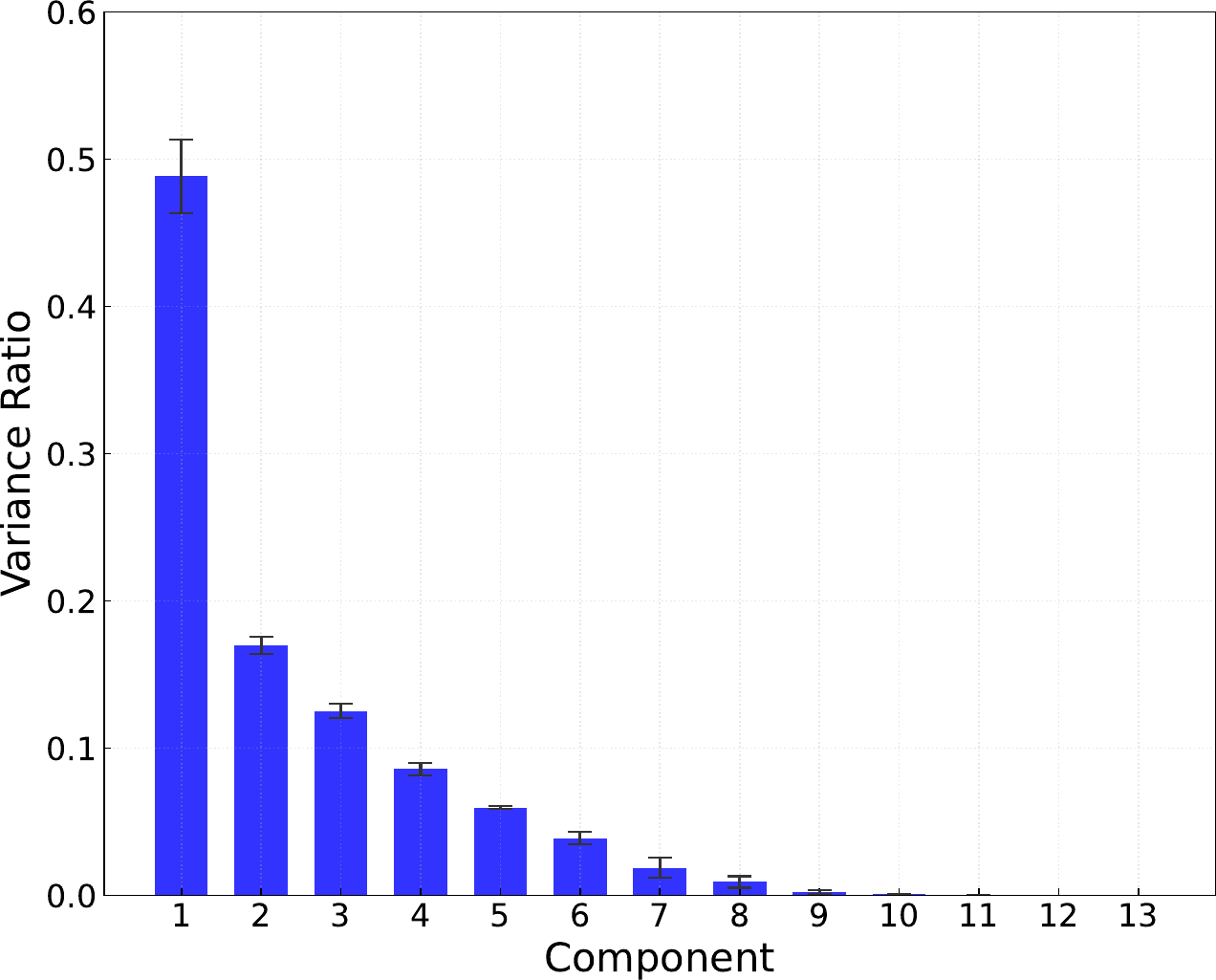}
% \hspace{0.5cm}
\includegraphics[scale=0.35]{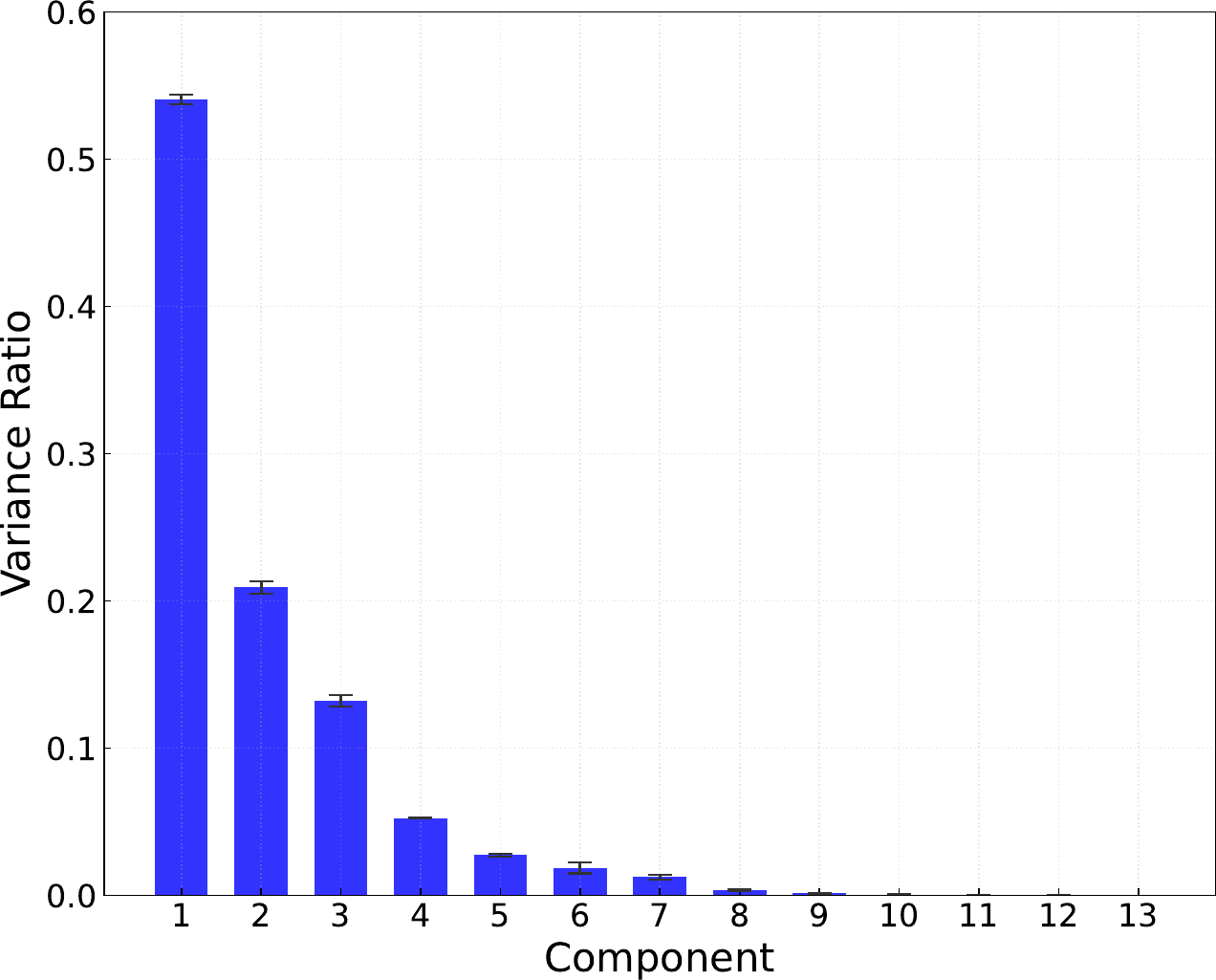}
% \hspace{0.5cm}
\caption{The {average} variance ratio of the training sets in Dataset A (left) and Dataset B (right).}
\label{fig_pca}
\end{figure}

Another method to study attributes is to project the data from the attribute space to a new space. In this new space, the direction with maximum variance is considered to contribute most to the data, while the directions with small variances can be removed without sacrificing crucial information. This method is called Principal Component Analysis (PCA) \citep{jolliffe2016principal}. {For each split dataset, we} normalize the samples and perform PCA on training sets. The variance ratio is {averaged and} depicted in Fig. \ref{fig_pca}. We find that {the most information of training data} is concentrated in the first 3 components (variance ratio larger than 0.1) both in Dataset A and B.

\subsection{Attribute Significance Estimator based on the Fractal Dimension (FDASE)}
\label{subsect_fdase}

Even though the RF model aggregates the attribute importance values from all the DTs (see Sec. \ref{subsect_attri_impo}), it may result in redundancy as it does not consider the correlations among important attributes. To address this issue, the FD theory is applied to estimate the potential contribution of each attribute to the dataset and measure the correlations among attributes \citep{belussi1995estimating,belussi1998self}. This method is based on the observation that independent attributes contribute more to the dataset, while correlated attributes contribute less.

Mathematically, for a dataset $\mathbf{A}$ with $E$ attributes: $\mathbb{A}=\{a_1,a_2,...,a_E\}$, the fractal dimension theory supposes that the potential existence of correlated attributes leads the set of points in the original $E$-dimensional space to describe one spatial object in a dimension that is lower than or equal to $E$. The dimension of the object represented by the dataset is called Intrinsic Dimension (ID), denoted by $D$, $D\in\mathbb{R}^+$. The ceiling of the ID $\lceil D \rceil$ is the minimum number of attributes that must be retained to keep the essential characteristics of a dataset \citep{deSousa2007}.

However, the ID of a dataset is difficult to obtain. Alternatively, we consider the ID projecting the dataset on a subset $\mathbf{C}\subset\mathbf{A}$, where $\mathbf{C}$ is defined by an attribute subspace $\mathbb{C}\subset\mathbb{A}$. It is named Partial Intrinsic Dimension (PID) on $\mathbb{C}$: $pD(\mathbb{C})$. Based on the definitions, the attribute $a_i\in(\mathbb{A}-\mathbb{C})$ increases $pD(\mathbb{C})$ by at most its Individual Contribution (IC) according to the degree of the correlation between $a_i$ and the attributes in $\mathbb{C}$. The IC of $a_i$, i.e., $iC(a_i)$, is the maximum potential contribution of the attribute $a_i$ to $pD(\mathbb{C})$. The greater the correlation between $a_i$ and the attributes in $\mathbb{C}$, the lower its contribution to $pD(\mathbb{C})$.

Also, $iC(a_i)$ can be measured by $pD(\{a_i\})$ and it ranges in $[0,1]$. A more independent distribution of the values of $a_i$ leads to $iC(a_i)$ closer to one, while a more structured distribution brings $iC(a_i)$ closer to zero \citep{deSousa2007}. Thus, the $E$-dimensional dataset can be seen as formed by adding the attributes with different contributions to the $D$-dimensional sub-dataset.

Moreover, the degree of correlations among attributes can be measured by a threshold $\xi$: a sub-dataset $\mathbf{B}\subset\mathbf{A}$ is said to be $\xi$-correlated to another sub-dataset $\mathbf{C}\subset\mathbf{A}$ (their attribute spaces $\mathbb{B}\cap\mathbb{C}=\emptyset$) if every attribute $a_i \in \mathbb{B}$ does not contribute more than $\xi*iC(a_i)$ to $pD(\mathbb{C})$. The threshold $\xi\in[0,1)$ tunes how strong the correlation between attributes in $\mathbb{B}$ and attributes in $\mathbb{C}$ should be to be detected.

\begin{figure}[htbp]
\centering
\includegraphics[scale=0.38]{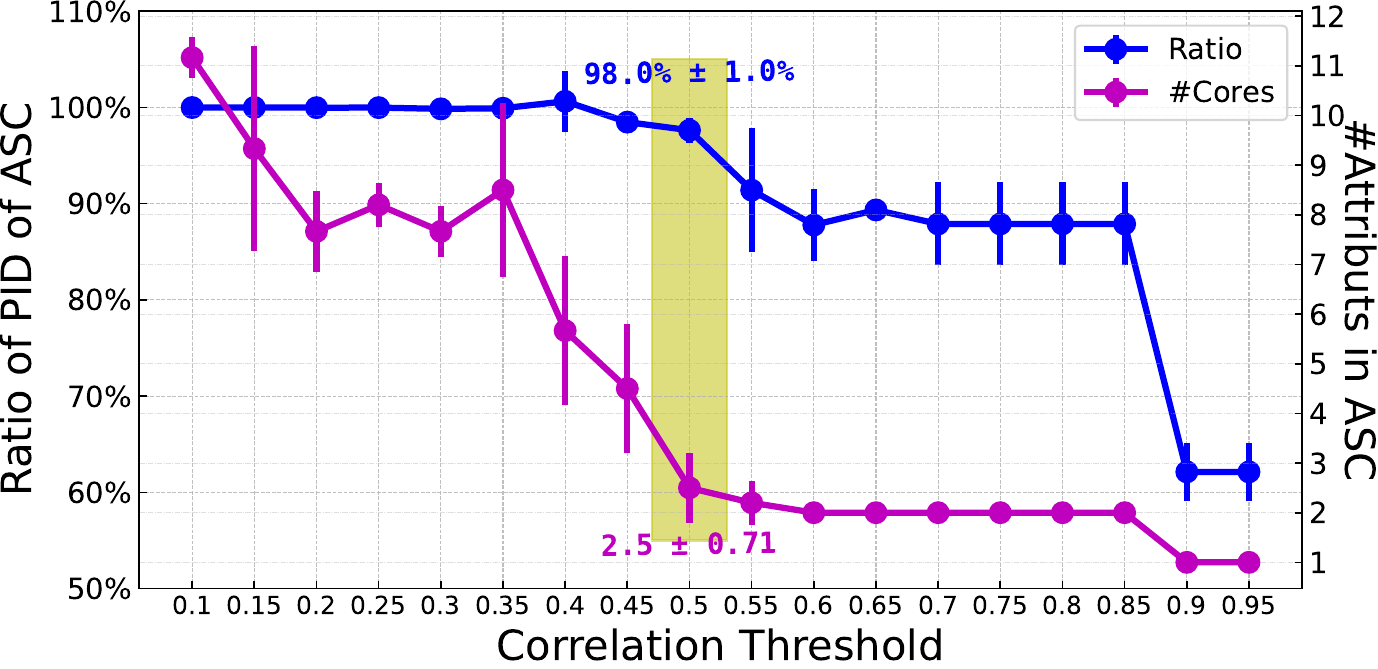}
\includegraphics[scale=0.38]{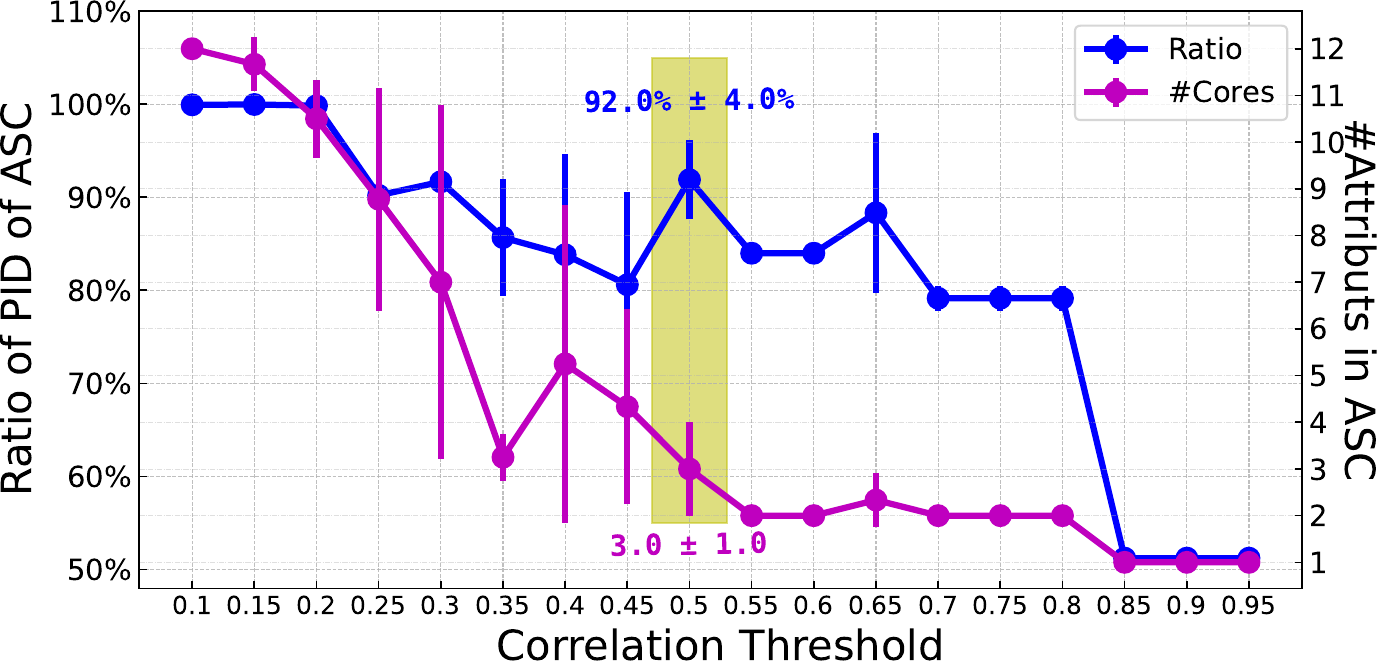}
\caption{The {average} ratio of the PID of ASC to ID {(magenta color)} and the {average} number of ASC {(blue color)} with different correlation threshold $\xi$ for the training sets of Dataset A (left) and Dataset B (right).}
\label{fig_fdase_ratio}
\end{figure}

\begin{figure}[htbp]
\centering
\includegraphics[scale=0.4]{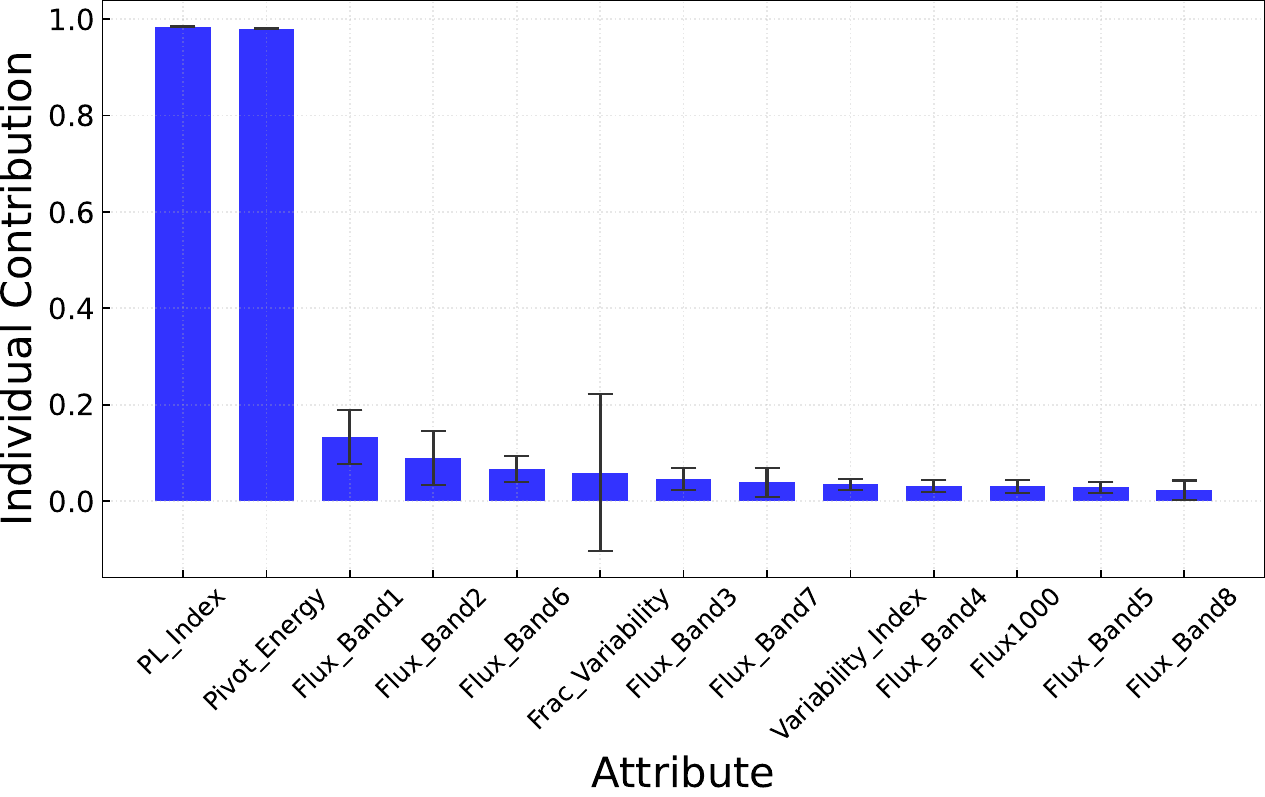}
\includegraphics[scale=0.4]{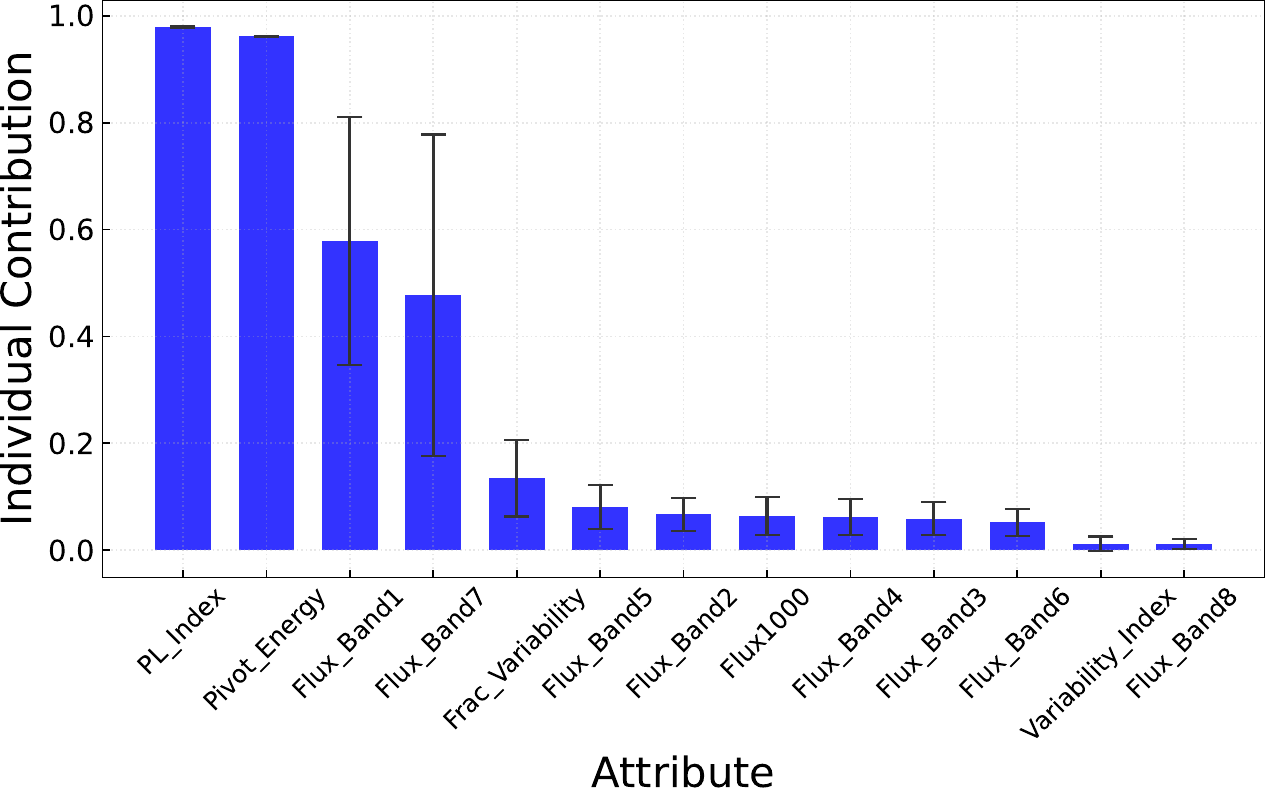}
\caption{{The IC for each attribute when $\xi=0.5$ for the training sets of Dataset A (left) and Dataset B (right).}}
\label{fig_fdase_ic}
\end{figure}

A greedy algorithm FDASE was developed by \citep{deSousa2007} to find a subset of attributes whose PID approaches the ID of the whole dataset. The result subset is called Attribute Set Core  $\xi C$ (ASC), with the given correlation threshold $\xi$ and scale range $n$. Therefore, the ratio $pD(\xi C)$/ID normalizes the contribution of ASC to the whole dataset. {Similarly, for each split dataset, we} utilize the FDASE algorithm on the training sets with scanning correlation threshold $\xi$ and compare the value of $pD(\xi C)$/ID with the number of ASC in Fig. \ref{fig_fdase_ratio}. We fix the scale range $n=50$ and plot the IC for each attribute in Fig. \ref{fig_fdase_ic} under a suitable $\xi$. From the figures, we find that when $\xi=0.5$ two attributes (`PL\_Index' and `Pivot\_Energy') contribute {around} $98\%$ information to Dataset A. While for Dataset B, {four attributes (`PL\_Index', `Pivot\_Energy', `Flux\_Band1' and `Flux\_Band7') contribute around $92\%$} information when {$\xi=0.5$}.

\section{Proposed Method}
\label{sect_proposed}

\subsection{Inverse Discrete Wavelet Transform}
\label{subsect_dwt}

The attribute analysis methods introduced in Sec. \ref{sect_data_prepro} are to find the important attributes or components that can be retained for the learning process, while other attributes or components are removed for dimension reduction. However, the crude removal may cause the loss of correlation features that degrade the performance.

We propose to retain all of the attributes but perform IDWT on the original samples for the representations at higher resolutions. After IDWT, the correlation features are supposed to be highlighted and meanwhile, the dimension is possibly reduced, which is important when dealing with big data.

The well-known wavelet transform may be the Discrete Wavelet Transform (DWT), which is usually implemented with filtering operations for high efficiency. The filters are designed based on the standpoint of multiresolution: the difference of information between the approximation of a signal at the resolutions $2^{m+1}$ and $2^m$ (where $m$ is an integer) can be extracted by decomposing this signal on an orthonormal basis of wavelets \citep{mallat1989theory}. The pyramidal structure of the wavelet filter bank makes it possible to infer the information at a low resolution from the information at a high resolution. IDWT is the converse process of DWT and provides the representations at high resolutions for DL.

However, in practical applications, a finite signal should be considered. The length of the signals varies at different resolutions, which is due to the operations of downsampling/upsampling and filtering. In the IDWT process, if denote $p$ and $s$ as the length of the signal at low resolution and high resolution, respectively, then \cite{rajmic2014discrete} gives:
\begin{equation}
\label{eqn_idwt_len}
s=\left\{
\begin{split}
&{2p-u+1},\ &\mathrm{for\ even}\ s+u-1,\\
&{2p-u+2},\ &\mathrm{for\ odd}\ s+u-1,
\end{split}\right.
\end{equation}
where $u$ is the length of the filters.

To perform IDWT on the original dataset, the attribute data of one sample will be seen as the signal at a lower resolution. As a result, the attributes should be rearranged in some order based on the wavelet theory. This can be achieved with the alignment of the information in the wavelet domain and attribute space, which is to be discussed next.

\subsection{Information in Wavelet Domain: $I_c$}
\label{subsect_wavelet_info}

\begin{figure}[htbp]
\centering
\includegraphics[scale=0.6]{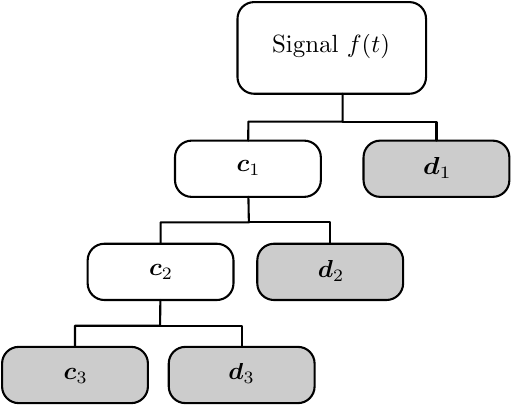}
\caption{Wavelet coefficients for a three-level pyramid transform.}
\label{fig_dwt_coefs}
\end{figure}

The process of DWT is to analyze the original signal from a fine scale to a coarse scale. The representation of a finite signal $f(t)$ in the wavelet domain after DWT is a collection of vectors:
\begin{equation}
\label{eqn_dwt_sequence}
    \{\boldsymbol{c}_J(k):k\in\mathbb{Z}\}\cup\{\boldsymbol{d}_j(k):1\le j \le J;k\in\mathbb{Z}\},
\end{equation}
where, $J\in \mathbb{N}$ is called decomposition level; $\boldsymbol{c}_J$ contains the approximation coefficients at level $J$, i.e., the lowest resolution, and $\boldsymbol{d}_j$ contains the detail coefficients at level $j$, i.e., the high resolutions.  An example of the three-level pyramid transform is illustrated in Fig. \ref{fig_dwt_coefs}.  We find that the original signal $f(t)$ can be seen as the approximation coefficients at the highest resolution, i.e., $\boldsymbol{c}_0$.

However, based on prior knowledge, it is often observed that at a specific level, the majority of information in a natural signal is typically presented in the approximation coefficients. Additionally, during the decomposition of the signal from level $j$ to $j+1$ in DWT, the information is typically reduced by half due to the downsampling operation. Hence, it can be found that the information of coefficients at different levels roughly respect to:
\begin{equation}
\label{eqn_dwt_info}
    I_c(\boldsymbol{d}_J)<\cdots<I_c(\boldsymbol{d}_2)<I_c(\boldsymbol{d}_{1})<I_c(\boldsymbol{c}_J).
\end{equation}

\begin{figure}[htbp]
\centering
\includegraphics[scale=0.2]{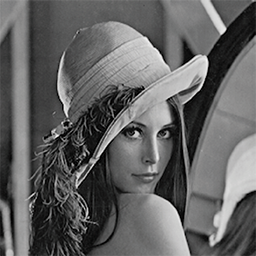}
(a)
% \hspace{0.5cm}
\includegraphics[scale=0.2]{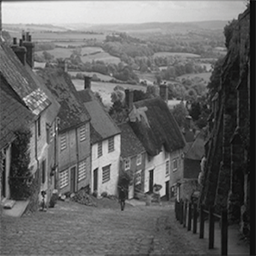}
(b)
% \hspace{0.5cm}
\includegraphics[scale=0.2]{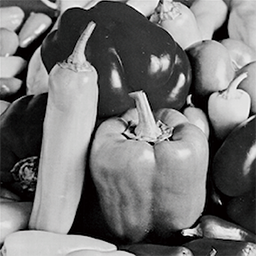}
(c)\\
% \hspace{0.5cm}\\[1cm]
\includegraphics[scale=0.30]{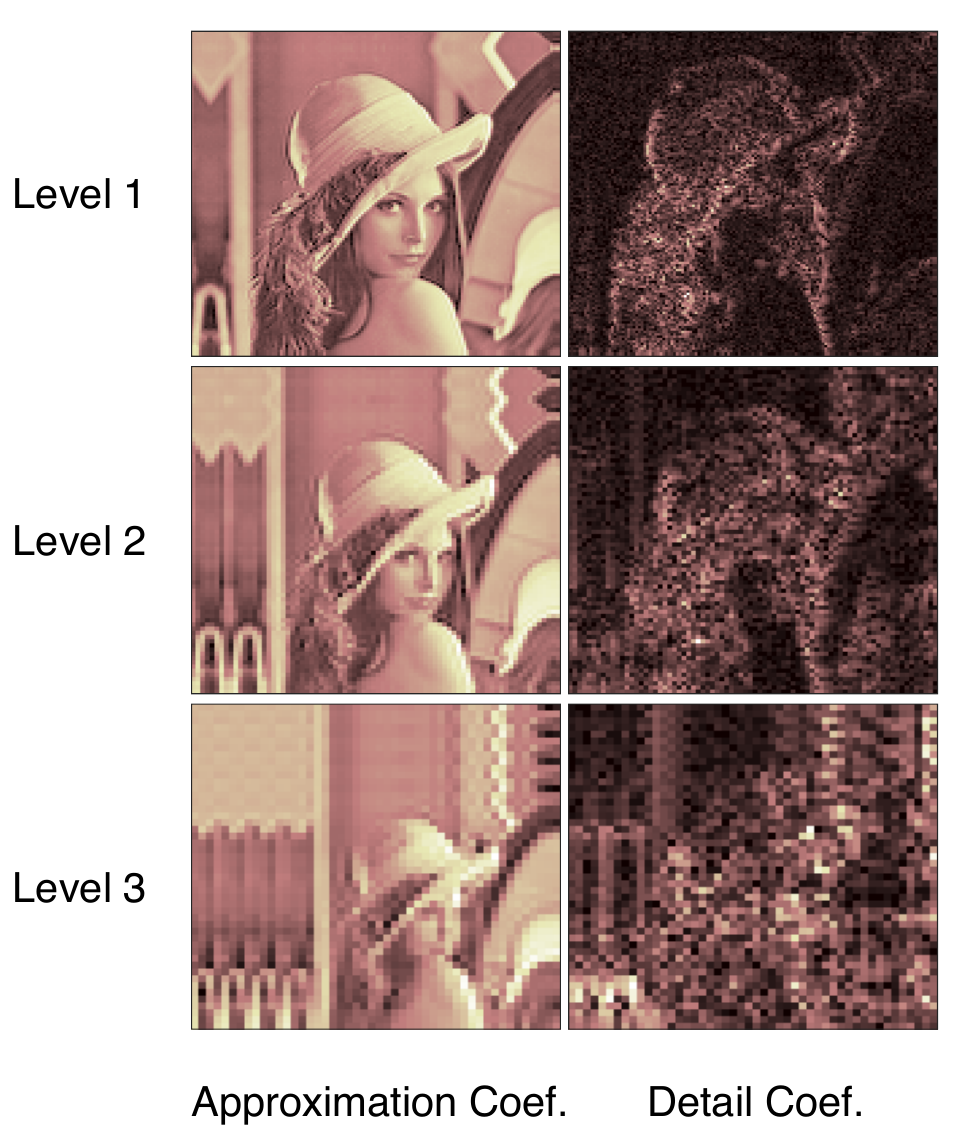}
(d)
% \hspace{0.5cm}
\includegraphics[scale=0.30]{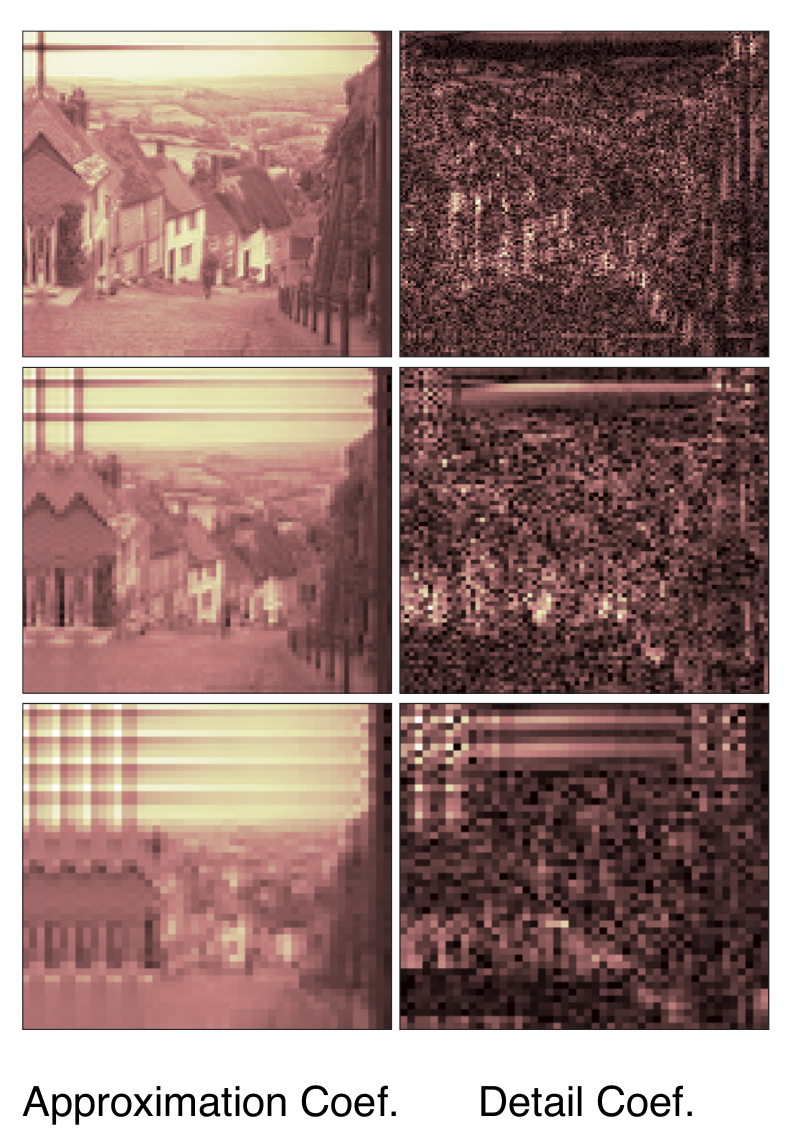}
(e)
% \hspace{0.5cm}
\includegraphics[scale=0.30]{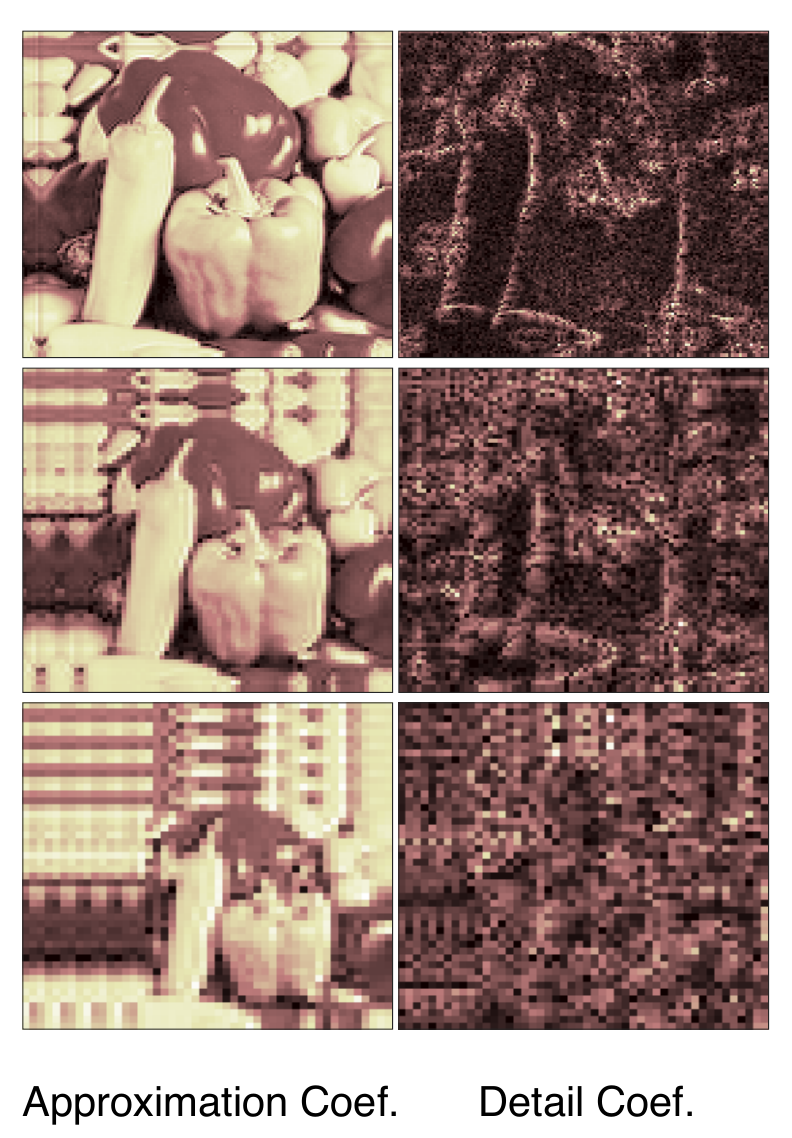}
(f)\\
% \hspace{0.5cm}\\[1cm]
\caption{The representations of images Lena, Goldhill, and Peppers in (a-c) spacial domain, and (d-f) wavelet domain, respectively. For display performance, the images of coefficients have been mapped into pink color. The detail coefficients at a particular level drawn in the figures are the sum of the corresponding horizontal, vertical, and diagonal detail coefficients.}
\label{fig_image_examp}
\end{figure}

As an example, the three-level DWT is performed on three images to intuitively illustrate the relationship, as shown in Fig. \ref{fig_image_examp}. These figures suggest that the approximation coefficients at each level retain the image contour and contain a significant amount of information, while the objects in the image cannot be easily identified through the detail coefficients. However, to some extent, the detail coefficients at a low level appear to provide more information than those at a high level, i.e., $I_c(\boldsymbol{d}_{j+1})<I_c(\boldsymbol{d}_{j})$.

To transform the coefficients from lower resolutions to higher resolutions in the IDWT process, the Wavelet Decomposition Vector (WDV) $\boldsymbol{c}$ and Bookkeeping Vector (BV) $\boldsymbol{l}$ are required according to multiresolution analysis. The WDV includes the coefficients shown in Eqn. \ref{eqn_dwt_sequence}:
\begin{equation}
\label{eqn_c}
\boldsymbol{c}=(\boldsymbol{c}_J,\boldsymbol{d}_J,\boldsymbol{d}_{J-1},\dots,\boldsymbol{d}_1),
\end{equation}
while the BV is made up of the number of coefficients in $\boldsymbol{c}$:
\begin{equation}
\label{eqn_l}
\boldsymbol{l}=(\mathrm{len}(\boldsymbol{c}_J),\mathrm{len}(\boldsymbol{d}_J),\mathrm{len}(\boldsymbol{d}_{J-1}),\dots,\mathrm{len}(\boldsymbol{d}_1), \mathrm{len}(f(t)))
\end{equation}
where $\mathrm{len}(\cdot)$ indicates the number of the elements in a vector.

\subsection{Information in Attribute Space: $I_a$}
\label{subsect_attri_info}

In the field of big data, an attribute's information reflects its contribution to the dataset. As explained in Sec. \ref{subsect_fdase}, the IC estimates the potential contribution of an individual attribute to the dataset. However, the presence of correlations among attributes means that the actual contribution of an attribute $a_i$, i.e., its information $I_a(\{a_i\})$, cannot be precisely determined if $a_i$ involves other attributes.

Moreover, as discussed, ASC is the smallest subset of attributes that can fully characterize the entire dataset. Given a correlation threshold $\xi$, the ASC $\xi C$ can be found out with the FDASE algorithm \citep{deSousa2007}. The attributes in ASC are not $\xi$-correlated with each other and they contain the most information in the dataset.

Based on the analysis, if $\mathbb{A}=\{a_1,a_2,\dots,a_E\}$ is denoted as the universal attribute set of the dataset that has $E$ attributes, then the real contribution of attribute $a_i\in (\mathbb{A}-\xi C)$ to the dataset can be seen as the degree of correlation between $a_i$ and the attributes in $\xi C$. The weaker the correlation between $a_i$ and $\xi C$, the higher {the} contribution of $a_i$ to the dataset, i.e., attribute $a_i$ contains more information. Therefore, the information of attribute $a_i$ can be estimated by
\begin{equation}
\label{eqn_info_attri}
I_a(\{a_i\})=| pD(\{a_i\}\cup \xi C)-pD(\xi C) |,\ a_i\in (\mathbb{A}-\xi C),
\end{equation}
where, $pD(\cdot)$ is the PID of a sub-dataset.

By estimating the information for each attribute $a_i\in (\mathbb{A}-\xi C)$, we can obtain a rough ordering of attributes based on their information. In this case, the data under the attributes can be treated as wavelet coefficients and arranged to form the WDV $\boldsymbol{c}$ and BV $\boldsymbol{l}$ for IDWT. The dataset is then transformed into representations at a higher resolution where the correlation features are presented. This is the main idea of our method, which will be detailed next.

\subsection{FDIDWT}
\label{subsect_fdidwt}

The $J$-level IDWT aims to transform the original data into the representations at the $J$-higher resolution. To achieve this, the WDV and BV must be constructed with the attribute data of each sample. However, the positions of the data in the WDV are determined by the estimation of the attribute information with Eqn. \ref{eqn_info_attri}. As a result, this approach is referred to as Fractal Dimension - Inverse Discrete Wavelet Transform (FDIDWT), or FDIDWT for short.

As discussed before, the attributes in ASC $\xi C\subset \mathbb{A}$ contains the most information of the dataset defined in $\mathbb{A}=\{a_1,a_2,\dots,a_E\}$. It is assumed that the information of attributes in $\mathbb{A}-\xi C$ respects to an ascending order:
\begin{equation}\footnotesize
\label{eqn_info_sort}
I_a(\{a_1\})<I_a(\{a_2\})<\cdots<I_a(\{a_i\})<\cdots<I_a(\{a_Q\})<I_a(\xi C),
\end{equation}
where, $a_i\in(\mathbb{A}-\xi C)$ and $1\le i\le Q$. If the number of attributes in $\xi C$ is $P$, then $P+Q=E$. Contrasting with the order of the information in the wavelet domain shown in Eqn. \ref{eqn_dwt_info}, we group the attributes in $\mathbb{A}-\xi C$ and obtain a similar order of the information in the attribute space:
\begin{equation}\small
\label{eqn_attribute_group}
I_a(D^M_J)<I_a(D^N_{J-1})<\cdots<I_a(D^L_{j})<\cdots<I_a(D^K_1)<I_a(\xi C),
\end{equation}
where, $D^L_j$ is the attribute group containing $L$ attributes whose information respects to Eqn. \ref{eqn_info_sort}, i.e.,
\begin{equation}
\begin{split}
&D^M_J=\{a_1,a_2,\dots,a_M\},\\
&D^N_{J-1} = \{a_{M+1},a_{M+2},\dots,a_{M+N}\},\\
&\dots\\
&D^K_1=\{a_{Q-K+1}, a_{Q-K+2},\dots,a_{Q}\},\\
&M+N+\cdots+K=Q.
\end{split}
\end{equation}

Thus the data under the attributes of $D^L_j$ and $\xi C$ can be taken as the detail coefficients $\boldsymbol{d}_j$ and approximation coefficients $\boldsymbol{c}_J$, respectively. Specifically, if denote $\boldsymbol{x}_j$ and $\boldsymbol{x}_{\xi C}$ as the data under the attributes in $D^L_j$ and $\xi C$ respectively for one sample of the original dataset,  the WDV $\boldsymbol{c}$ and BV $\boldsymbol{l}$ for the IDWT process can be constructed as:
\begin{equation}
\label{eqn_vectors}
\begin{split}
&\boldsymbol{c}=\left(\boldsymbol{x}_{\xi C},\boldsymbol{x}_J,\boldsymbol{x}_{J-1},\dots,\boldsymbol{x}_1\right),\\
&\boldsymbol{l}=\left(P,M,N,\dots,K,O\right),
\end{split}
\end{equation}
where $O$ is the number of attributes in the transformed dataset, i.e., the \textit{transformed dimension} in the new attribute space, which can be derived with Eqn. \ref{eqn_idwt_len}. Tab. \ref{tab_fdidwt} shows the procedure of the proposed FDIDWT method. To better describe how the algorithm works, we fix the decomposition level $J=3$. Moreover, as in the common case, the length of filters $u$ for the pyramid transform is set to be even. Then:
\begin{table*}[hbtp]\small
\centering
\caption{The proposed FDIDWT method.}
\label{tab_fdidwt}
\begin{tabular*}{\hsize}{rl}
\toprule
\midrule
\textbf{Input:} & original dataset defined on attribute space $\mathbb{A}$ \\
& decomposition level $J$ \\
% & transformed dimension $O$ \\
& correlation threshold $\xi$ \\
& scale range $n$\\
\textbf{Output:} & new dataset defined on attribute space $\mathbb{B}$ \\\midrule
1: & run FDASE algorithm to find out the ASC $\xi C$, denote its length as $P$; \\
2: & calculate $I_a({a_i})$, $a_i\in(\mathbb{A}-\xi C)$, with Eqn. \ref{eqn_info_attri}; \\
3: & arrange the attributes of $\mathbb{A}$ into $\mathbb{A}'$ according to the order of $I_a({a_i})$ as shown in Eqn. {\ref{eqn_info_sort}}; \\
4: & based on $P$, calculate the length of other groups, insert placeholder attributes if needed;\\
5: & construct the WDV $\boldsymbol{c}$ and BV $\boldsymbol{l}$;\\
6: & perform IDWT on each normalized sample with $\boldsymbol{c}$ and $\boldsymbol{l}$; \\
7: & output the new dataset defined on $\mathbb{B}$; \\ 
\bottomrule
\end{tabular*}
\end{table*}

\begin{enumerate}
\item Step 1-3: Given the correlation threshold $\xi$ and scale range $n$, the \textit{FDASE} algorithm \citep{deSousa2007} is used to find out the attribute set core $\xi C$. Then the information $I_a(\{a_i\})$ of the remaining attributes $a_i\in(\mathbb{A}-\xi C)$ are calculated. The attributes of $\mathbb{A}$ are arranged into a set $\mathbb{A}'$ according to the ascending order of $I_a(\{a_i\})$\footnote{Here, $\mathbb{A}'$ is a sequence since the information of its attributes respects to an order.}. For instance, if the relationship in Eqn. \ref{eqn_info_sort} holds, then
\begin{equation}
\label{eqn_sort_attri}
\mathbb{A}'=\xi C \cup \{a_1,a_2,\dots,a_i,\dots,a_Q\}.
\end{equation}

In our experiment, we scanned the correlation threshold $\xi$ in the range [0.05, 1.0] with step 0.05, and set scale range $n=50$. We got the ASC \{$a_3$, $a_1$\} and \{$a_3$, $a_1$, $a_5$\} for Dataset A and B, respectively, as shown in Fig. \ref{fig_fdase_ratio} and \ref{fig_fdase_ic}. Then the arranged attribute set $\mathbb{A}'$ for the two datasets will be \{$a_3$, $a_1$, $a_5$, $a_{13}$, $a_{10}$, $a_2$, $a_9$, $a_{12}$, $a_4$, $a_8$, $a_{11}$, $a_7$, $a_6$\} and \{$a_3$, $a_1$, $a_5$, $a_{13}$, $a_4$, $a_{12}$, $a_{11}$, $a_7$, $a_8$, $a_6$, $a_9$, $a_2$, $a_{10}$\}.

\item Step 4-5: $P$ is the number of attributes in $\xi C$. In the proposed method, the length of the approximation coefficients at level $J$ is assumed to be not shorter than $P$. If this length equals $P$, then Fig. \ref{fig_groups} shows other groups divided from $\mathbb{A}'$. $C_3^P$ denotes the approximation coefficient group of attributes in $\xi C$. Note that, the original dimension $E$, i.e., the number of attributes of the original dataset, may be smaller than the length of WDV $\boldsymbol{c}$ due to the given decomposition level $J$ and filter length $u$. In this case, some necessary \textit{placeholder attributes} "$*$" are inserted following the ASC group $C_3^P$. The role of the placeholder attributes is to ensure IDWT works while not changing the information of data. Therefore, the values under the placeholder attributes are set to zeros so that they contribute nothing to the dataset.
\begin{figure*}[htbp]
\centering
\includegraphics[scale=0.6]{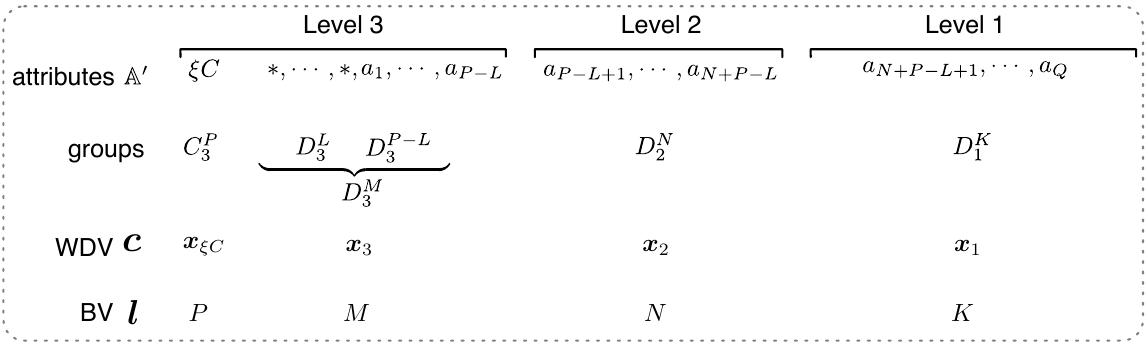}
\caption{The groups divided from sorted attributes. Taking decomposition level $J=3$ as an example.}
\label{fig_groups}
\end{figure*}

In the signal processing field, signals are always transformed into the wavelet domain with DWT for processing, and the lengths of the wavelet coefficients at each level are recorded for the reconstruction with IDWT. However, there is no DWT in the proposed method. Hence, many cases exist for the lengths of the divided groups based on Eqn. \ref{eqn_idwt_len}:
\begin{equation}
\label{eqn_length}
\begin{split}
&M=P,\\
&N=\left\{
\begin{split}
&{2M-u+1},\ \ \ \mathrm{for\ even}\ N+u-1\\
&{2M-u+2},\ \ \ \mathrm{for\ odd}\ N+u-1
\end{split}\right.,\\
&K=\left\{
\begin{split}
&{2N-u+1},\ \ \ \mathrm{for\ even}\ K+u-1\\
&{2N-u+2},\ \ \ \mathrm{for\ odd}\ K+u-1
\end{split}\right..
\end{split}
\end{equation}
The length of the placeholder attributes $L$ can be derived based on the equality of the length of WDV, i.e., $P+M+N+K=E+L$. Then the WDV and BV are constituted as follows:
\begin{equation}
\label{eqn_vectors_three}
\begin{split}
&\boldsymbol{c}=\left(\boldsymbol{x}_{\xi C},\boldsymbol{x}_3,\boldsymbol{x}_{2},\boldsymbol{x}_1\right),\\
&\boldsymbol{l}=\left(P,M,N,K,O\right),
\end{split}
\end{equation}
where $O$ is the transformed dimension computed with:
\begin{equation}
O=\left\{
\begin{split}
&{2K-u+1},&\ \mathrm{for\ even}\ O+u-1\\
&{2K-u+2},&\ \mathrm{for\ odd}\ O+u-1
\end{split}\right..
\end{equation}
We used the Daubechies filter which can be indicated by its length $u$. For example, the filter length $u = 8$ indicates that the "db4" wavelet filter was used. Under the assumption of the max decomposition level $J=3$, we built the WDV and BV for two missions in Tab. \ref{tab_vectors}.

\begin{table*}[htbp]
\scriptsize
\centering
\caption{The wavelet decomposition vector and bookkeeping vector for the IDWT in two missions.}
\label{tab_vectors}
\begin{tabular}{ccccc}
\toprule
\midrule
\textbf{Decomposition Level} & \textbf{WDV} & \textbf{BV} & \textbf{Wavelet Name} & \textbf{Case} \\ \midrule
\multirow{13}{*}{1} &\multirow{13}{*}{\begin{tabular}{l} A: (\{$x_3,x_1,x_5,x_{13},x_{10},x_2,x_9$\}, \{$0,x_{12},x_4,x_8,x_{11},x_7,x_6$\}) \\ B: (\{$x_3$,$x_1$,$x_5$,$x_{13}$,$x_4$,$x_{12}$,$x_{11}$\}, \{0,$x_7$,$x_8$,$x_6$,$x_9$,$x_2$,$x_{10}$\}) \end{tabular}}&(7,7,13)& db1 & \#1 \\
&&(7,7,12)& db2 & \#2 \\
&&(7,7,11)& db2 & \#3 \\
&&(7,7,10)& db3 & \#4 \\
&&(7,7,9)& db3 & \#5 \\
&&(7,7,8)& db4 & \#6 \\
&&(7,7,7)& db4 & \#7 \\
&&(7,7,6)& db5 & \#8 \\
&&(7,7,5)& db5 & \#9 \\
&&(7,7,4)& db6 & \#10 \\
&&(7,7,3)& db6 & \#11 \\
&&(7,7,2)& db7 & \#12 \\
&&(7,7,1)& db7 & \#13 \\ \cmidrule{1-5}
\multirow{2}{*}{3} &\multirow{2}{*}{\begin{tabular}{l} A: (\{$x_3,x_1,x_5$\},\{$x_{13},x_{10},x_2$\},\{$x_9, x_{12},x_4$\},\{$x_8,x_{11},x_7,x_6$\}) \\ B: (\{$x_3$,$x_1$,$x_5$\},\{$x_{13}$,$x_4$,$x_{12}$\},\{$x_{11}$,$x_7$,$x_8$\},\{$x_6$,$x_9$,$x_2$,$x_{10}$\}) \end{tabular}}&(3,3,3,4,5)& db2 & \#14 \\
&&(3,3,3,4,6)& db2 & \#15 \\ \bottomrule 
\multicolumn{5}{l}{Value 0 in the WDV means one placeholder attribute is inserted.}
\end{tabular}
\end{table*}

\item Step 6-7: Finally, IDWT is performed on each sample of the normalized dataset, and a new dataset defined on $\mathbb{B}=\{b_1,b_2,\dots,b_i,\dots,b_O\}$ is generated. $b_i$ is the $i$th new attribute.
\end{enumerate}

\subsection{A Lightweight CNN}
\label{subsect_nn}
\begin{figure}[htbp]
\centering
\includegraphics[scale=0.3]{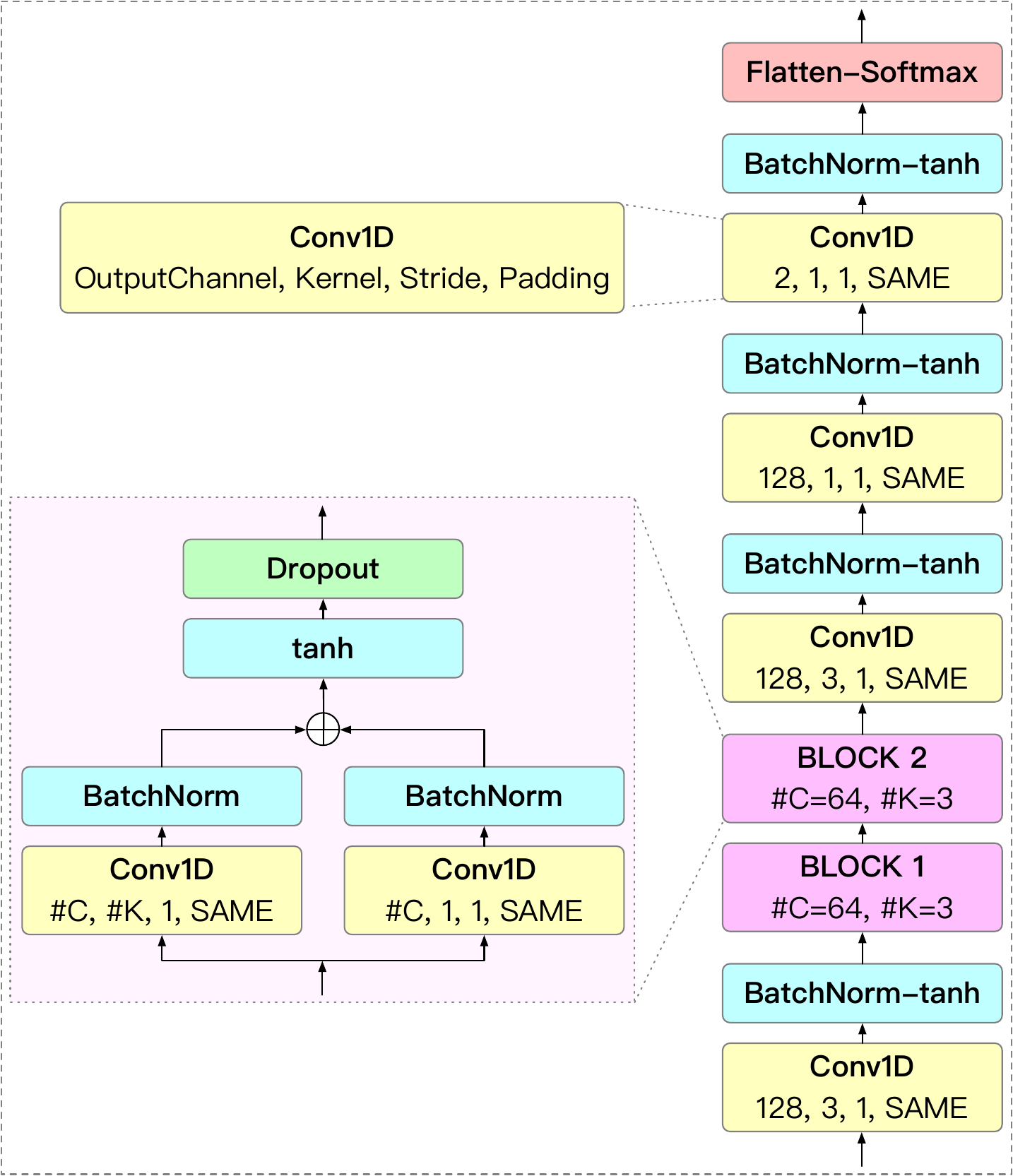}
\caption{The proposed lightweight MatchboxConv1D model.}
\label{fig_model}
\end{figure}

Since the new attribute space $\mathbb{B}$ is considered to respect some natural order after FDIDWT, we believe the convolution operation with a kernel will learn more features and achieve better performance. The kernel of CNN determines a visual field on the adjacent attributes in DL. Thus we utilized a lightweight CNN model modified from the Matchbox net \citep{matchbox} but with a smaller size of only 94k parameters for high efficiency. We refer it to as "MatchboxConv1D" and its structure is depicted in Fig. \ref{fig_model}.

\section{Experiments and Results}
\label{sect_exps}
\begin{figure}[htbp]
\centering
\includegraphics[scale=0.4]{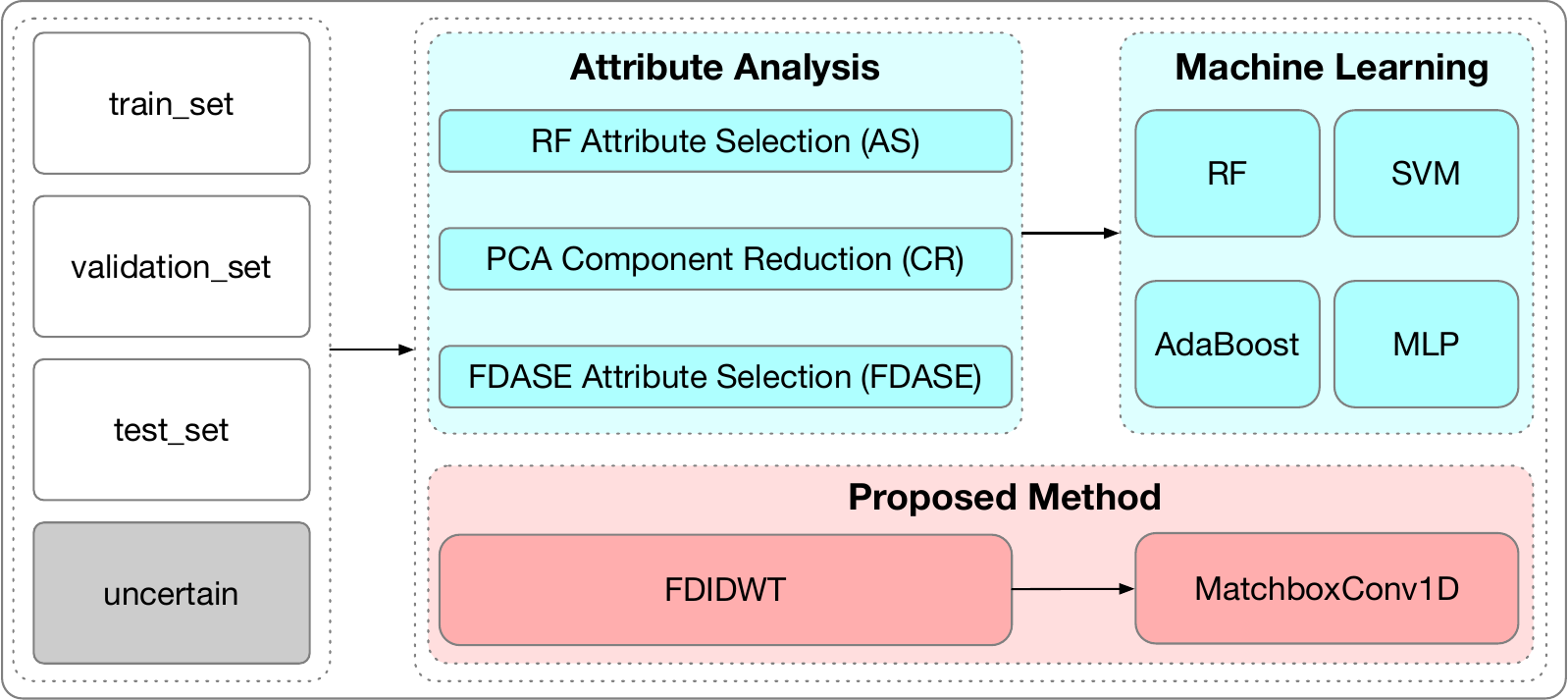}
\caption{The flowchart of attribute analysis and learning process.}
\label{fig_process}
\end{figure}

With the 3 attribute analysis methods introduced in Sec. \ref{sect_data_prepro}, we independently used 4 ML classifiers, i.e, RF, Support Vector Machine (SVM), AdaBoost, and Multilayer Perceptron (MLP) to carry out the Mission A and Mission B. The hyper-parameters are listed in Tab. \ref{tab_hyperparameters_ML}. While the FDIDWT method is designed to work with the MatchboxConv1D model proposed in Sec. \ref{subsect_nn}, and the hyper-parameters are shown in Tab. \ref{tab_hyperparameters_proposed}. After training the models with the training set, fine-tuning with the validation set, and evaluating with the test set, the uncertain samples, i.e., the uncertain sources in Mission A and the BCUs in Mission B, are finally predicted with the model achieving the highest test accuracy. This flowchart is illustrated in Fig. \ref{fig_process}. 

{We performed the experiments on each split dataset, and the accuracy results were averaged over 10 test sets. In the experiments, we extensively evaluated the performance of an increasing number of the most important attributes or principal components. These attributes and components were sorted in descending order of importance and variance ratio, respectively, as shown in Fig. \ref{fig_attri_importance}, Fig. \ref{fig_pca}, and Fig. \ref{fig_fdase_ic}. The ML methods were implemented with scikit-learn \citep{pedregosa2011scikit}, and the training process of DNNs was accelerated by TensorFlow \citep{abadi2016tensorflow} on NVIDIA GeForce RTX 3060. The code used in this experiment has been uploaded to GitHub\footnote{\url{https://github.com/chtfrank/astro_FDIDWT}}.}

\begin{table*}[htbp]
\scriptsize
\centering
\caption{The hyper-parameters of attribute analysis methods and classifiers.}
\label{tab_hyperparameters_ML}
% \begin{threeparttable}
\begin{tabular}{rlrl}
\toprule
\midrule
\textbf{Attribute Analysis Methods} & \textbf{Hyper-parameters} & \textbf{Classifiers} & \textbf{Hyper-parameters} \\ 
\midrule

RF Attribute Selection {(AS)} & \begin{tabular}[c]{@{}l@{}} \#DTs: 50000\\ criterion: entropy \end{tabular} & RF & \begin{tabular}[c]{@{}l@{}} \#DTs: 700\\criterion: entropy \end{tabular} \\
\midrule

PCA Component Reduction {(CR)} & \begin{tabular}[c]{@{}l@{}} normalization: StandardScaler \end{tabular} & SVM & \begin{tabular}[c]{@{}l@{}} normalization: StandardScaler\\ C = 5.0\\degree = 1 \end{tabular} \\
\midrule

FDASE Attribute Selection {(FDASE)} & \begin{tabular}[c]{@{}l@{}} scale range: 50\\ correlation threshold: 0.5 \end{tabular} & {AdaBoost} & \begin{tabular}[c]{@{}l@{}} normalization: MinMaxScaler \\ \#estimators: 300\\ learning rate: 0.5 \end{tabular} \\ \midrule

 & & MLP & \begin{tabular}[c]{@{}l@{}} hidden units: [32,32,64,32,32]\\ dropout: 0.5\\ activation function: Leaky ReLU \\ batch size: 64\\ epochs: 500\\ learning rate: 0.0005 \end{tabular} \\

\bottomrule
% \multicolumn{4}{l}{(A) and (B) means the Dataset A and B, respectively.} \\
\end{tabular}%
% \end{threeparttable}
\end{table*}

\begin{table*}[htbp]
\scriptsize
\centering
\caption{The hyper-parameters of proposed method and model.}
\label{tab_hyperparameters_proposed}
% \begin{threeparttable}
\begin{tabular}{rlrl}
\toprule
\midrule
\textbf{Proposed Method} & \textbf{Hyper-parameters} & \textbf{Proposed Model} & \textbf{Hyper-parameters} \\ 
\midrule

FDIDWT & \begin{tabular}[c]{@{}l@{}} wavelet type: Daubechies\\max decomposition level: 3 \end{tabular} & MatchboxConv1D & \begin{tabular}[c]{@{}l@{}} dropout: 0.9\\ activation function: tanh \\ batch size: 64\\ epochs: 500\\ learning rate: 0.0005\\more can be found in Fig. \ref{fig_model} \end{tabular} \\

\bottomrule
\end{tabular}%
% \end{threeparttable}
\end{table*}

\begin{figure}[htbp]
\centering
\includegraphics[scale=0.5]{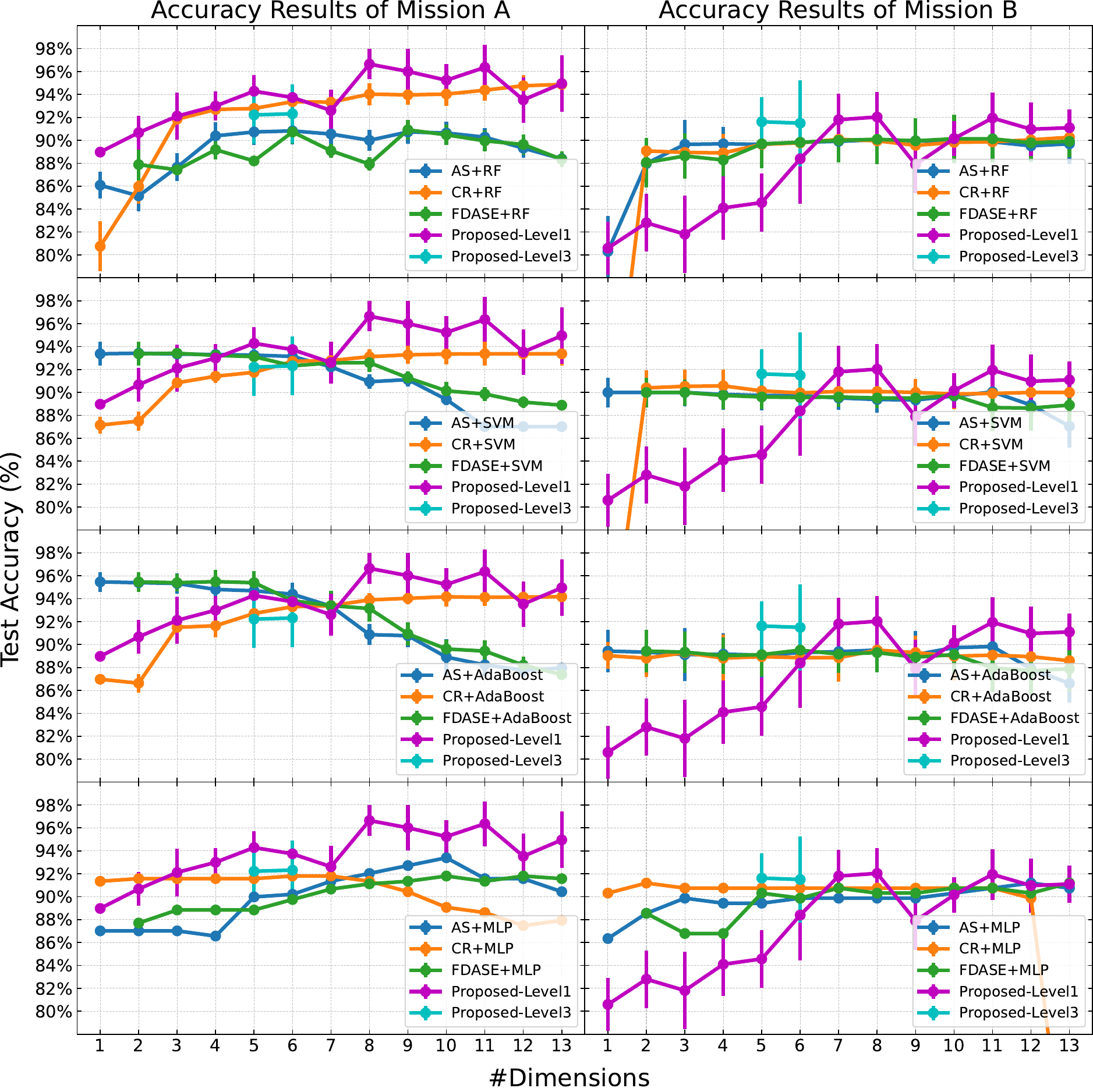}
\caption{{The comparison of average test accuracy results in two missions. The results of the proposed methods in each mission are depicted in every panel.}}
\label{fig_test_acc_results}
\end{figure}

The {average} accuracy results of the commonly used attribute analysis methods and classifiers on test sets are compared in Tab. \ref{tab_accuracy_A} and \ref{tab_accuracy_B} for two missions. {These results are also illustrated in Fig. \ref{fig_test_acc_results}. From the results, we find that the highest test accuracy is {$95.49\% \pm 1.05\%$} in mission A, which comes from the  AdaBoost classifier working with the FDASE attribute selection method in the case of reducing two dimensions. In mission B, the highest test accuracy is {$91.19\% \pm 0.0\%$} which results from the MLP classifier working with different attribute analysis methods but reduces at most one dimension.}

\begin{table*}[htbp]
\scriptsize
\centering
% \begin{threeparttable}
\caption{{The average test accuracy results of Mission A with the ML methods.}}
\label{tab_accuracy_A}
\begin{tabular}{cccccc}
\toprule
\midrule
\textbf{Attribute Analysis Methods} & \textbf{\#Dimension} & \textbf{RF} & \textbf{SVM} & \textbf{AdaBoost} & \textbf{MLP} \\ \midrule

\multirow{13}{*}{RF Attribute Selection (AS)} & 1 & $86.08\%\pm1.15\%$  & $87.02\%\pm0.0\%$  & $87.95\%\pm0.41\%$ & $87.02\%\pm0.0\%$ \\
& 2 & $85.15\%\pm1.34\%$  & $87.02\%\pm0.0\%$  & $87.77\%\pm0.61\%$ & $87.02\%\pm0.0\%$ \\
& 3 & $87.65\%\pm1.21\%$  & $87.02\%\pm0.0\%$  & $88.22\%\pm1.0\%$ & $87.02\%\pm0.0\%$ \\
& 4 & $90.39\%\pm1.18\%$  & $89.38\%\pm0.49\%$  & $88.91\%\pm1.58\%$ & $86.56\%\pm0.0\%$ \\
& 5 & $90.73\%\pm1.16\%$  & $91.12\%\pm0.59\%$  & $90.77\%\pm1.0\%$ & $89.98\%\pm0.0\%$ \\
& 6 & \hllightgray{$\mathbf{90.82\%\pm1.18\%}$}  & $90.93\%\pm0.64\%$  & $90.87\%\pm0.91\%$ & $90.2\%\pm0.0\%$ \\
& 7 & $90.55\%\pm0.98\%$  & $92.23\%\pm0.88\%$  & $93.35\%\pm0.93\%$ & $91.34\%\pm0.0\%$ \\
& 8 & $90.0\%\pm0.92\%$  & $93.14\%\pm0.81\%$  & $94.4\%\pm1.03\%$ & $92.03\%\pm0.0\%$ \\
& 9 & $90.73\%\pm1.03\%$  & $93.26\%\pm0.78\%$  & $94.69\%\pm1.02\%$ & $92.71\%\pm0.0\%$ \\
& 10 & $90.64\%\pm0.97\%$  & $93.3\%\pm0.83\%$  & $94.81\%\pm1.14\%$ & \hllightgray{$\mathbf{93.39\%\pm0.0\%}$} \\
& 11 & $90.27\%\pm0.79\%$  & $93.35\%\pm0.78\%$  & $95.33\%\pm0.88\%$ & $91.57\%\pm0.0\%$ \\
& 12 & $89.32\%\pm0.83\%$  & \hllightgray{$\mathbf{93.42\%\pm0.76\%}$}  & $95.4\%\pm0.77\%$ & $91.57\%\pm0.0\%$ \\
& 13 & $88.38\%\pm0.59\%$  & $93.37\%\pm1.04\%$  & \hllightgray{$\mathbf{95.47\%\pm0.86\%}$} & $90.43\%\pm0.0\%$ \\ \midrule

\multirow{13}{*}{PCA Component Reduction (CR)} & 1 & $80.75\%\pm2.17\%$  & $87.15\%\pm0.75\%$  & $86.97\%\pm0.14\%$ & $87.93\%\pm0.0\%$ \\
& 2 & $85.97\%\pm1.5\%$  & $87.49\%\pm0.81\%$  & $86.63\%\pm0.82\%$ & $87.47\%\pm0.0\%$ \\
& 3 & $91.85\%\pm1.05\%$  & $90.84\%\pm0.53\%$  & $91.5\%\pm1.16\%$ & $88.61\%\pm0.0\%$ \\
& 4 & $92.69\%\pm0.75\%$  & $91.41\%\pm0.61\%$  & $91.64\%\pm1.0\%$ & $89.07\%\pm0.0\%$ \\
& 5 & $92.78\%\pm0.77\%$  & $91.75\%\pm0.64\%$  & $92.71\%\pm1.08\%$ & $90.43\%\pm0.0\%$ \\
& 6 & $93.37\%\pm0.61\%$  & $92.71\%\pm0.6\%$  & $93.3\%\pm0.88\%$ & $91.34\%\pm0.0\%$ \\
& 7 & $93.33\%\pm0.95\%$  & $92.78\%\pm0.59\%$  & $93.39\%\pm0.9\%$ & \hllightgray{$\mathbf{91.8\%\pm0.0\%}$} \\
& 8 & $94.03\%\pm0.97\%$  & $93.12\%\pm0.67\%$  & $93.9\%\pm0.58\%$ & $91.8\%\pm0.0\%$ \\
& 9 & $93.96\%\pm0.89\%$  & $93.28\%\pm0.78\%$  & $94.03\%\pm0.52\%$ & $91.57\%\pm0.0\%$ \\
& 10 & $94.03\%\pm1.05\%$  & $93.35\%\pm0.9\%$  & \hllightgray{$\mathbf{94.17\%\pm0.83\%}$} & $91.57\%\pm0.0\%$ \\
& 11 & $94.37\%\pm0.89\%$  & \hllightgray{$\mathbf{93.37\%\pm1.04\%}$}  & $94.12\%\pm0.74\%$ & $91.57\%\pm0.0\%$ \\
& 12 & $94.76\%\pm0.91\%$  & $93.37\%\pm1.04\%$  & $94.15\%\pm1.04\%$ & $91.57\%\pm0.0\%$ \\
& 13 & \hllightgray{$\mathbf{94.87\%\pm0.93\%}$}  & $93.37\%\pm1.04\%$  & $94.17\%\pm0.85\%$ & $91.34\%\pm0.0\%$ \\ \midrule

\multirow{12}{*}{FDASE Attribute Selection (FDASE)} & 2 & $87.88\%\pm1.36\%$  & $88.88\%\pm0.45\%$  & $87.38\%\pm0.37\%$ & $87.7\%\pm0.0\%$ \\
& 3 & $87.43\%\pm0.45\%$  & $89.16\%\pm0.49\%$  & $88.2\%\pm0.76\%$ & $88.84\%\pm0.0\%$ \\
& 4 & $89.18\%\pm0.85\%$  & $89.86\%\pm0.61\%$  & $89.43\%\pm0.95\%$ & $88.84\%\pm0.0\%$ \\
& 5 & $88.2\%\pm0.34\%$  & $90.14\%\pm0.79\%$  & $89.61\%\pm0.8\%$ & $88.84\%\pm0.0\%$ \\
& 6 & $90.73\%\pm1.05\%$  & $91.28\%\pm0.57\%$  & $90.93\%\pm1.03\%$ & $89.75\%\pm0.0\%$ \\
& 7 & $89.09\%\pm0.61\%$  & $92.6\%\pm0.8\%$  & $93.14\%\pm1.12\%$ & $90.66\%\pm0.0\%$ \\
& 8 & $87.93\%\pm0.59\%$  & $92.57\%\pm0.75\%$  & $93.39\%\pm1.27\%$ & $91.12\%\pm0.0\%$ \\
& 9 & \hllightgray{$\mathbf{90.91\%\pm0.87\%}$}  & $92.32\%\pm0.61\%$  & $93.83\%\pm0.89\%$ & $91.34\%\pm0.0\%$ \\
& 10 & $90.48\%\pm0.88\%$  & $93.12\%\pm0.69\%$  & $95.4\%\pm1.02\%$ & \hllightgray{$\mathbf{91.8\%\pm0.0\%}$} \\
& 11 & $89.95\%\pm0.93\%$  & $93.23\%\pm0.79\%$  & \hllightgray{$\mathbf{95.49\%\pm1.05\%}$} & $91.34\%\pm0.0\%$ \\
& 12 & $89.64\%\pm0.89\%$  & \hllightgray{$\mathbf{93.42\%\pm0.76\%}$}  & $95.4\%\pm0.77\%$ & $91.8\%\pm0.0\%$ \\
& 13 & $88.34\%\pm0.69\%$  & $93.37\%\pm1.04\%$  & $95.47\%\pm0.86\%$ & $91.57\%\pm0.0\%$ \\ \bottomrule
\multicolumn{6}{l}{The highest test accuracy for each combination of the attribute analysis method and the classifier is highlighted by a gray color.} \\
\end{tabular}
\end{table*}

\begin{table*}[htbp]
\scriptsize
\centering
% \begin{threeparttable}
\caption{{The average test accuracy results of Mission B with the ML methods.}}
\label{tab_accuracy_B}
\begin{tabular}{cccccc}
\toprule
\midrule
\textbf{Attribute Analysis Methods} & \textbf{\#Dimension} & \textbf{RF} & \textbf{SVM} & \textbf{AdaBoost} & \textbf{MLP} \\ \midrule

\multirow{13}{*}{RF Attribute Selection (AS)} & 1 & $80.31\%\pm3.08\%$  & $87.05\%\pm1.85\%$  & $86.61\%\pm1.69\%$ & $86.34\%\pm0.0\%$ \\
& 2 & $87.97\%\pm1.97\%$  & $88.9\%\pm1.67\%$  & $87.89\%\pm1.97\%$ & $88.55\%\pm0.0\%$ \\
& 3 & $89.65\%\pm2.12\%$  & \hllightgray{$\mathbf{90.04\%\pm0.89\%}$}  & \hllightgray{$\mathbf{89.82\%\pm1.45\%}$} & $89.87\%\pm0.0\%$ \\
& 4 & $89.69\%\pm1.46\%$  & $89.69\%\pm1.02\%$  & $89.74\%\pm1.86\%$ & $89.43\%\pm0.0\%$ \\
& 5 & $89.65\%\pm1.79\%$  & $89.34\%\pm1.05\%$  & $89.12\%\pm2.08\%$ & $89.43\%\pm0.0\%$ \\
& 6 & $89.82\%\pm1.93\%$  & $89.38\%\pm1.13\%$  & $89.52\%\pm1.83\%$ & $89.87\%\pm0.0\%$ \\
& 7 & $89.91\%\pm1.55\%$  & $89.52\%\pm1.03\%$  & $89.38\%\pm2.19\%$ & $89.87\%\pm0.0\%$ \\
& 8 & \hllightgray{$\mathbf{90.04\%\pm1.62\%}$}  & $89.74\%\pm0.95\%$  & $89.3\%\pm2.04\%$ & $89.87\%\pm0.0\%$ \\
& 9 & $89.96\%\pm1.54\%$  & $89.74\%\pm1.3\%$  & $89.07\%\pm2.32\%$ & $89.87\%\pm0.0\%$ \\
& 10 & $89.91\%\pm1.69\%$  & $89.82\%\pm1.35\%$  & $89.16\%\pm1.82\%$ & $90.31\%\pm0.0\%$ \\
& 11 & $89.87\%\pm1.38\%$  & $90.0\%\pm1.21\%$  & $89.12\%\pm2.29\%$ & $90.75\%\pm0.0\%$ \\
& 12 & $89.52\%\pm1.39\%$  & $90.0\%\pm1.14\%$  & $89.34\%\pm1.81\%$ & \hllightgray{$\mathbf{91.19\%\pm0.0\%}$} \\
& 13 & $89.69\%\pm1.78\%$  & $90.0\%\pm1.3\%$  & $89.43\%\pm1.83\%$ & $90.75\%\pm0.0\%$ \\ \midrule

\multirow{13}{*}{PCA Component Reduction (CR)} & 1 & $61.15\%\pm2.41\%$  & $65.42\%\pm1.85\%$  & $89.03\%\pm1.16\%$ & $63.44\%\pm0.0\%$ \\
& 2 & $89.07\%\pm1.11\%$  & $90.4\%\pm1.54\%$  & $88.81\%\pm1.65\%$ & $89.87\%\pm0.0\%$ \\
& 3 & $88.94\%\pm1.4\%$  & $90.53\%\pm1.44\%$  & $89.3\%\pm1.54\%$ & $90.75\%\pm0.0\%$ \\
& 4 & $88.9\%\pm1.66\%$  & \hllightgray{$\mathbf{90.57\%\pm1.44\%}$}  & $88.81\%\pm2.08\%$ & $90.75\%\pm0.0\%$ \\
& 5 & $89.65\%\pm1.64\%$  & $90.13\%\pm1.27\%$  & $88.94\%\pm1.72\%$ & $90.75\%\pm0.0\%$ \\
& 6 & $89.78\%\pm1.47\%$  & $89.96\%\pm1.33\%$  & $88.85\%\pm1.64\%$ & $90.75\%\pm0.0\%$ \\
& 7 & $90.09\%\pm1.43\%$  & $90.09\%\pm1.25\%$  & $88.85\%\pm2.08\%$ & $90.75\%\pm0.0\%$ \\
& 8 & $89.96\%\pm1.58\%$  & $90.09\%\pm1.32\%$  & \hllightgray{$\mathbf{89.52\%\pm1.73\%}$} & $90.75\%\pm0.0\%$ \\
& 9 & $89.56\%\pm1.47\%$  & $90.0\%\pm1.18\%$  & $89.3\%\pm1.48\%$ & $90.75\%\pm0.0\%$ \\
& 10 & $89.82\%\pm1.46\%$  & $89.87\%\pm1.31\%$  & $88.99\%\pm2.13\%$ & $90.75\%\pm0.0\%$ \\
& 11 & $89.87\%\pm1.66\%$  & $89.91\%\pm1.38\%$  & $89.08\%\pm2.02\%$ & $90.75\%\pm0.0\%$ \\
& 12 & $90.04\%\pm1.6\%$  & $90.0\%\pm1.3\%$  & $88.94\%\pm1.67\%$ & \hllightgray{$\mathbf{91.19\%\pm0.0\%}$} \\
& 13 & \hllightgray{$\mathbf{90.26\%\pm1.74\%}$}  & $90.0\%\pm1.3\%$  & $88.59\%\pm1.56\%$ & $90.31\%\pm0.0\%$ \\ \midrule

\multirow{12}{*}{FDASE Attribute Selection (FDASE)} & 2 & $88.06\%\pm2.15\%$  & $88.9\%\pm1.67\%$  & $87.89\%\pm1.97\%$ & $88.55\%\pm0.0\%$ \\
& 3 & $88.63\%\pm1.99\%$  & $88.63\%\pm1.97\%$  & $87.71\%\pm2.13\%$ & $86.78\%\pm0.0\%$ \\
& 4 & $88.28\%\pm1.78\%$  & $88.68\%\pm2.0\%$  & $87.97\%\pm2.13\%$ & $86.78\%\pm0.0\%$ \\
& 5 & $89.69\%\pm2.14\%$  & $89.74\%\pm0.95\%$  & $89.12\%\pm1.62\%$ & $90.31\%\pm0.0\%$ \\
& 6 & $89.82\%\pm2.08\%$  & $89.52\%\pm0.74\%$  & $88.9\%\pm1.41\%$ & $89.87\%\pm0.0\%$ \\
& 7 & $90.0\%\pm1.92\%$  & $89.52\%\pm0.9\%$  & $89.3\%\pm1.74\%$ & $90.75\%\pm0.0\%$ \\
& 8 & $90.09\%\pm2.16\%$  & $89.6\%\pm0.86\%$  & $89.16\%\pm1.64\%$ & $90.31\%\pm0.0\%$ \\
& 9 & $89.96\%\pm1.99\%$  & $89.56\%\pm0.91\%$  & \hllightgray{$\mathbf{89.52\%\pm1.8\%}$} & $90.31\%\pm0.0\%$ \\
& 10 & \hllightgray{$\mathbf{90.13\%\pm2.09\%}$}  & $89.6\%\pm1.06\%$  & $89.12\%\pm1.9\%$ & $90.75\%\pm0.0\%$ \\
& 11 & $90.13\%\pm2.06\%$  & $89.74\%\pm1.18\%$  & $89.03\%\pm1.6\%$ & $90.75\%\pm0.0\%$ \\
& 12 & $89.78\%\pm1.61\%$  & \hllightgray{$\mathbf{90.0\%\pm1.14\%}$}  & $89.34\%\pm1.81\%$ & $90.31\%\pm0.0\%$ \\
& 13 & $89.91\%\pm1.37\%$  & $90.0\%\pm1.3\%$  & $89.43\%\pm1.83\%$ & \hllightgray{$\mathbf{91.19\%\pm0.0\%}$} \\ \bottomrule
\multicolumn{6}{l}{The highest test accuracy for each combination of the attribute analysis method and the classifier is highlighted by a gray color.} \\
\end{tabular}
\end{table*}

Moreover, Tab. \ref{tab_accuracy_new_method} shows the {average} test accuracy results of the proposed method with different transformed dimensions (see Tab. \ref{tab_vectors}) for two missions. {We find that case \#6 achieves the highest test accuracy and outperforms the results of Tab. \ref{tab_accuracy_A} and Tab. \ref{tab_accuracy_B}, meanwhile reducing five dimensions.}
Finally, the prediction results of uncertain sources and BCUs in two missions are listed in Tab. \ref{tab_prediction}.

\begin{table*}[htbp]
\scriptsize
\centering
\caption{The {average} test accuracy results of two missions with the FDIDWT method and MatchboxConv1D model.}
\label{tab_accuracy_new_method}
\begin{tabular}{ccccc}
\toprule
\midrule
% \textbf{Transformed Dimension} & \textbf{Case} & \textbf{Test Accuracy} \\ \midrule
\multirow{2}{*}{\textbf{\#Dimension}} & \multirow{2}{*}{\textbf{Case}} & \multicolumn{2}{c}{\textbf{Test Accuracy}} \\ \cmidrule{3-4}
&& \textbf{Mission A} & \textbf{Mission B} \\ \midrule
13 & \#1 & $94.95\%\pm2.47\%$ & $91.1\%\pm1.61\%$ \\
12 & \#2 & $93.53\%\pm1.98\%$ & $90.97\%\pm2.35\%$ \\
11 & \#3 & $96.36\%\pm1.95\%$ & $91.94\%\pm2.21\%$ \\
10 & \#4 & $95.24\%\pm1.44\%$ & $90.16\%\pm1.54\%$ \\
9 & \#5 & $96.01\%\pm1.96\%$ & $87.89\%\pm2.53\%$ \\
8 & \#6 & \hllightgray{$\mathbf{96.65\%\pm1.32\%}$} & \hllightgray{$\mathbf{92.03\%\pm2.2\%}$} \\
7 & \#7 & $92.61\%\pm1.81\%$ & $91.8\%\pm2.28\%$ \\
6 & \#8 & $93.74\%\pm0.99\%$ & $88.39\%\pm3.94\%$ \\
5 & \#9 & $94.29\%\pm1.4\%$ & $84.57\%\pm2.53\%$ \\
4 & \#10 & $93.0\%\pm1.25\%$ & $84.1\%\pm2.78\%$ \\
3 & \#11 & $92.11\%\pm2.07\%$ & $81.81\%\pm3.38\%$ \\
2 & \#12 & $90.67\%\pm1.47\%$ & $82.8\%\pm2.51\%$ \\
1 & \#13 & $88.97\%\pm0.01\%$ & $80.59\%\pm2.31\%$ \\
5 & \#14 & $92.22\%\pm2.51\%$ & $91.62\%\pm2.17\%$ \\
6 & \#15 & $92.31\%\pm2.57\%$ & $91.50\%\pm3.76\%$ \\

\bottomrule
\end{tabular}
\end{table*}

% \begin{figure}[htbp]
% \centering
% \includegraphics[scale=0.4]{cm_A.png}
% \includegraphics[scale=0.4]{cm_B.png}
% \caption{The confusion matrices in case \#4 for the test datasets of Dataset A (left) and Dataset B (right), respectively.}
% \label{fig_confusion}
% \end{figure}

\begin{table*}[htbp]
\scriptsize
\centering
\caption{The prediction results of uncertain sources in Mission A and BCUs in Mission B.}
\label{tab_prediction}
\begin{tabular}{ccccc}
\toprule
\midrule
\textbf{Mission} & \textbf{Source name} & \textbf{Predict Probability} & \textbf{Predict Class}\\
\midrule
\multirow{6}{*}{A} & 4FGL J0000.3-7355  & $0.9963 \pm 0.0028$ & AGN  \\
& 4FGL J0000.5+0743  & $0.9948 \pm 0.0048$   & AGN           \\
& 4FGL J0000.7+2530  & $0.9822 \pm 0.0229$   & AGN           \\
& 4FGL J0001.6+3503  & $0.9805 \pm 0.0153$   & AGN           \\
& 4FGL J0002.1+6721c & $0.9698 \pm 0.0111$   & Non-AGN       \\
& .... & .... & .... \\
\midrule
\multirow{6}{*}{B} & 4FGL J0001.2+4741 & $0.9315 \pm 0.0404$  & BL Lac (BCU)\\
& 4FGL J0001.6-4156 & $0.9887 \pm 0.0159$  & BL Lac \\
& 4FGL J0001.8-2153 & $0.8635 \pm 0.0966$  & BL Lac (BCU) \\
& 4FGL J0002.3-0815 & $0.9881 \pm 0.0047$  & BL Lac \\
& 4FGL J0002.4-5156 & $0.8595 \pm 0.1072$  & BL Lac (BCU) \\
& .... & .... & .... \\
\bottomrule
\multicolumn{4}{l}{The full version of this table is available online.}
\end{tabular}
\end{table*}

\section{Discussion}
\label{sect_discussion}
\subsection{The Proposed Method}
The extraction of correlation features among the attributes of \textit{Fermi} sources is a promising research field. 
The commonly used attribute analysis methods such as the estimation of attribute importance (see Sec. \ref{subsect_attri_impo}) or principal components (see Sec. \ref{subsect_pca}) aim to find the significant attributes or components, and remove the unimportant ones for dimension reduction. 
However, the correlation features will be removed with these methods, which results in inferior classification performance (see the results of Tab. \ref{tab_accuracy_A} and Tab. \ref{tab_accuracy_B}).

The FDASE gives a global perspective of the whole dataset and estimates the correlation features among all of the attributes based on fractal dimension theory. 
The ASC is the resulting attribute subset of the FDASE algorithm and its data contains most of the information of the dataset under some correlation threshold and scale range. 
% These two hyper-parameters must be set in advance but we have empiric values. 
Nevertheless, the retention of ASC also suffers from the loss of correlation features (see the results of FDASE attribute selection in Tab. \ref{tab_accuracy_A} and Tab. \ref{tab_accuracy_B}). 
Thus, we propose to rearrange the original attributes for highlighting correlation features at higher resolutions based on FDASE and IDWT, which is referred to as the proposed FDIDWT method.

Additionally, the resulting attribute space is considered to respect some natural order after FDIDWT. 
We believe that the convolution operation will extract more correlation features from the ordered attributes. 
Meanwhile, the structure of the CNN gives more potentiality for better classification performance. 
Therefore, by combining FDIDWT with the MatchboxConv1D model, we obtain the best test accuracy for both Mission A and B in case {\#6}. 
Besides, the dimension is transferred to be {8}, which is significant for reducing the computing burden in big astronomical data processing. 
It may also be concluded that most features of Dataset A and B exist at one higher resolution since case {\#6} corresponds to decomposition level 1. (see Tab. \ref{tab_vectors}).

\begin{figure}
\centering
\includegraphics[scale=0.6]{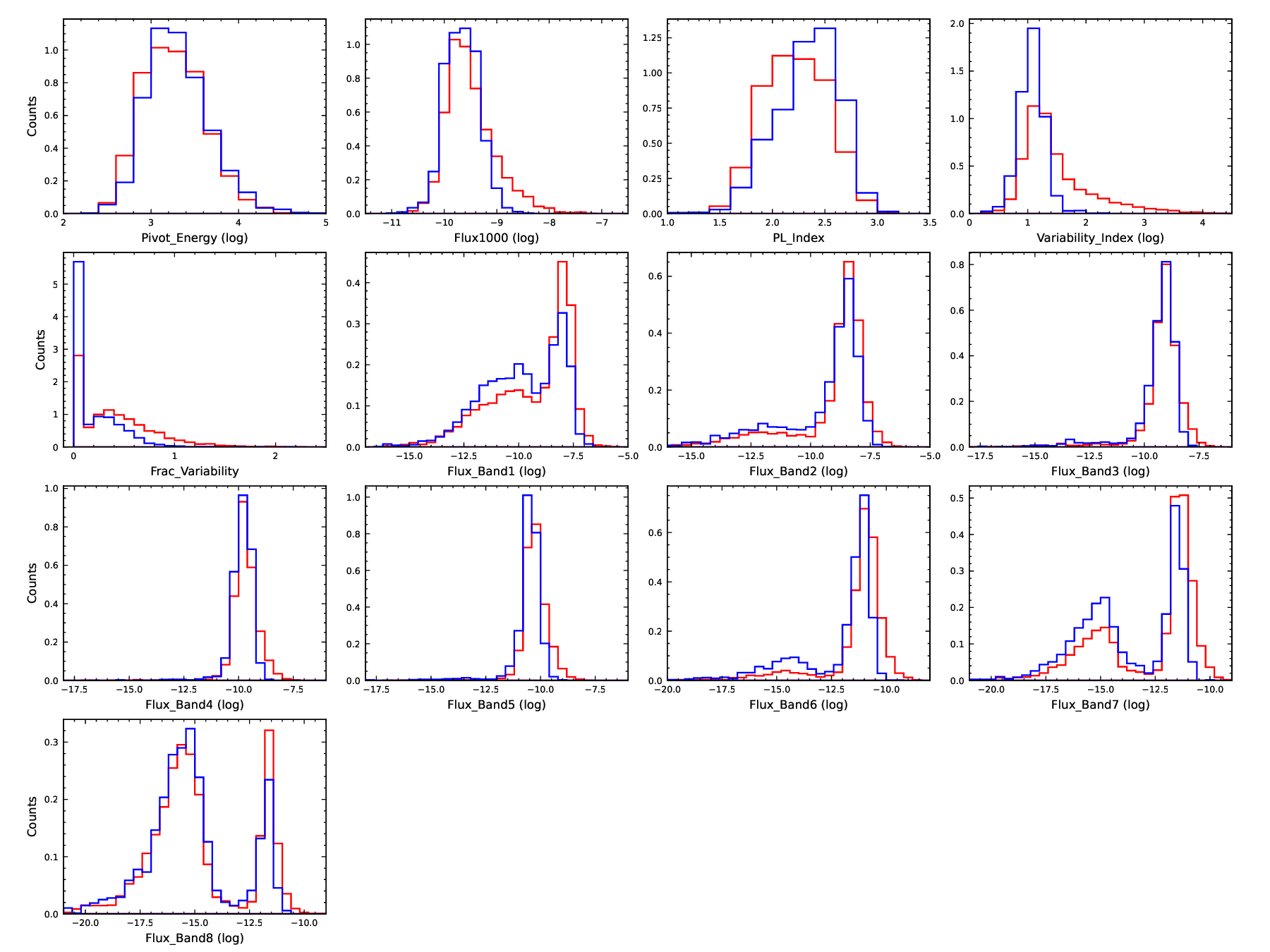}
\caption{The comparison of 13 attributes between 4FGL\_DR3 AGNs and predicted AGNs.
The histogram of 4FGL\_DR3 AGNs is shown in red and the histogram of predicted AGNs is shown in blue, the area under the histogram integrates into one.}
\label{dis_ma}
\end{figure}

\begin{figure}
\centering
\includegraphics[scale=0.6]{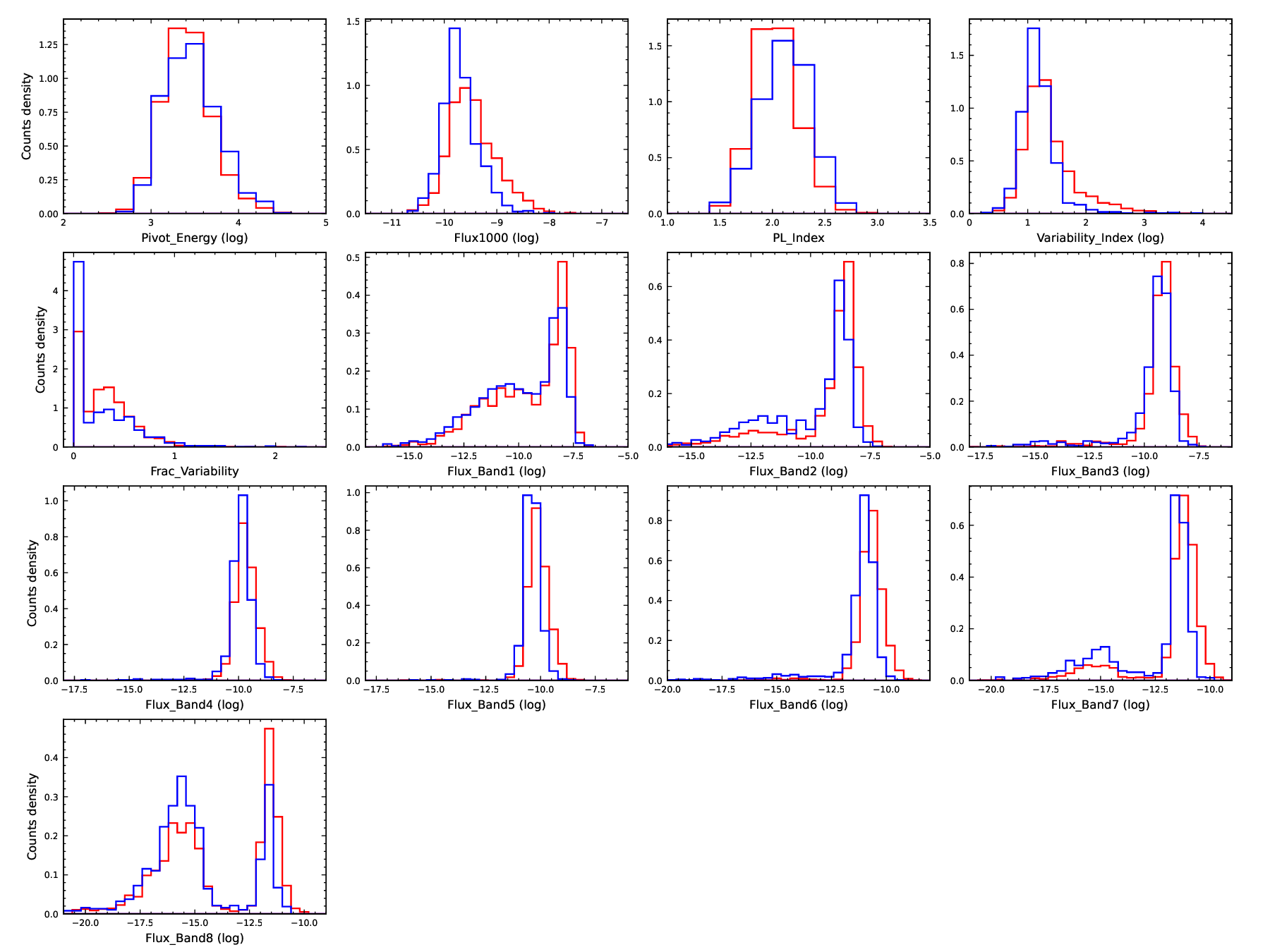}
\caption{The comparison of 13 attributes between 4FGL\_DR3 BL Lacs and predicted BL Lacs.
The histogram of 4FGL\_DR3 BL Lacs is shown in red and the histogram of predicted BL Lacs is shown in blue, the area under the histogram integrates into one.}
\label{dis_mbb}
\end{figure}

\begin{figure}
\centering
\includegraphics[scale=0.6]{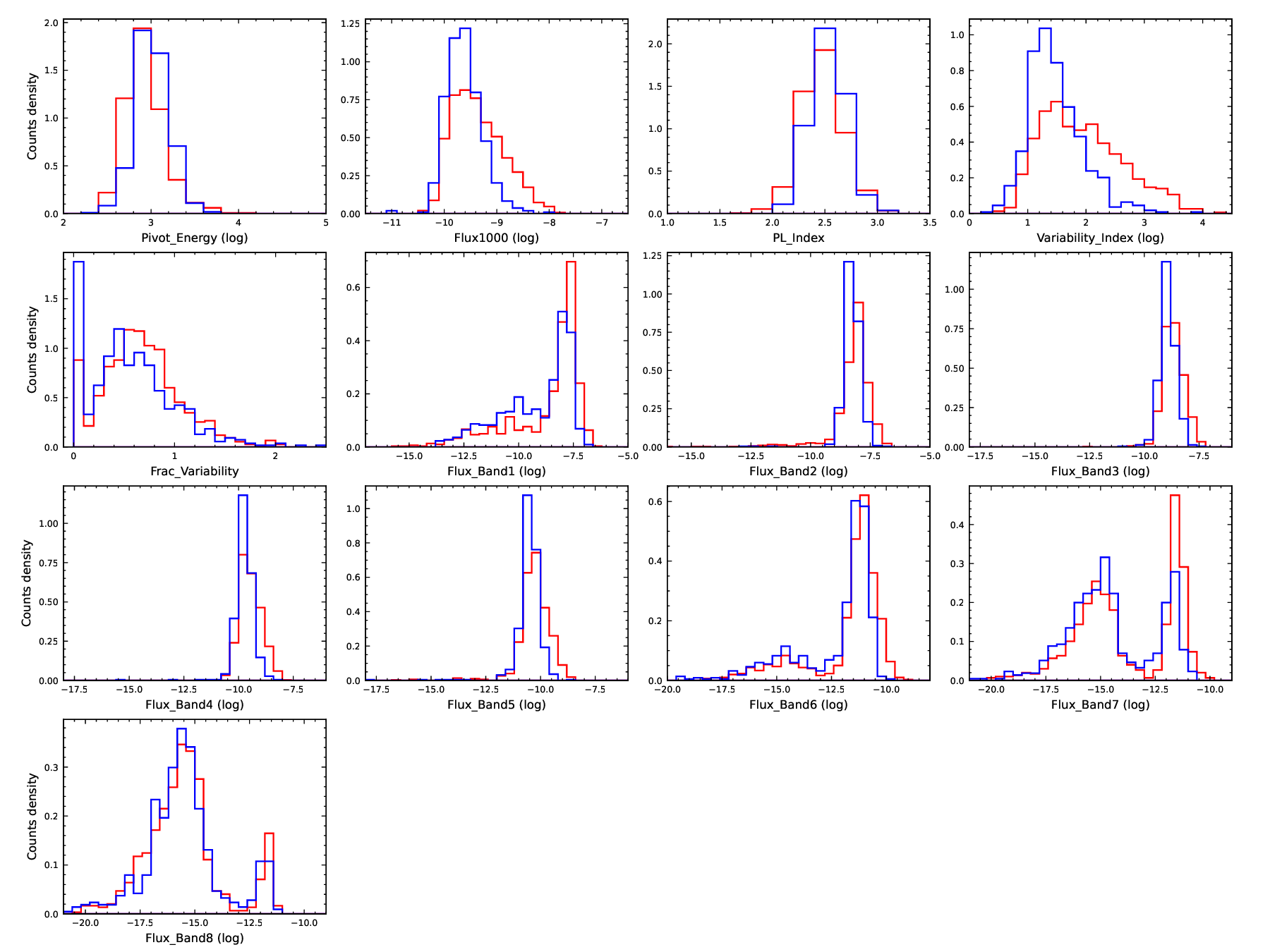}
\caption{The comparison of 13 attributes between 4FGL\_DR3 FSRQs and predicted FSRQs.
The histogram of 4FGL\_DR3 FSRQs is shown in red and the histogram of predicted FSRQs is shown in blue, the area under the histogram integrates to one.}
\label{dis_mbf}
\end{figure}

The prediction results have been carried out with an accuracy of {$96.65\% \pm 1.32\%$} and {$92.03\% \pm 2.2\%$} for Mission A and Mission B, respectively, using the proposed method, and the results are listed in Tab. \ref{tab_prediction}.
A general comparison between the predicted AGNs of Mission A and the originally confirmed AGNs is shown in Fig. \ref{dis_ma}, and the comparison between the predicted BL Lacs of Mission B and the originally confirmed BL Lacs is shown in Fig. \ref{dis_mbb}, and the comparison between the predicted FSRQs of Mission B and the originally confirmed FSRQs is shown in Fig. \ref{dis_mbf}.
From the comparison, we can find that the distribution shapes of the 13 attributes of the predicted sources resemble the shapes of the original 4FGL\_DR3 sources with the corresponding classification, in general. The results indicate that our predicted sources are correctly classified.

Moreover, we notice that the histograms of the attribute `Variability\_Index' show a longer high-variation-index tail for the original 4FGL\_DR3 sources than for those predicted sources (including predicted AGNs, predicted BL Lacs, and predicted FSRQs in Fig. \ref{dis_ma}-\ref{dis_mbf}, respectively), and the predicted sources have more contribution on the high-variation-index head than the original 4FGL\_DR3 ones.
This result suggests that our method has the advantage of finding less variable sources from those uncertain sources.
Also, we notice that the original 4FGL\_DR3 sources give more contribution to the histogram tails than the predicted sources while the predicted sources give more contribution to the histogram head for the attributes of multi-band intensities (`Flux1000', `Flux\_Band1', `Flux\_Band2', `Flux\_Band3', `Flux\_Band4', `Flux\_Band5', `Flux\_Band6', `Flux\_Band7', and `Flux\_Band8').
It suggests that our method has the advantage of finding relatively faint $\gamma$-ray sources from those uncertain ones.
It also encourages that our method should be able to make efforts to identify less variable and faint sources in the era of survey telescopes, e.g., the Large Synoptic Survey Telescope (LSST, \citealp{LSST2009}), China Space Station Telescope (CSST, \citealp{Zhan2011}), etc.

\subsection{The Classification Results of Mission A and Mission B}
\begin{figure}
\centering
\includegraphics[scale=0.6]{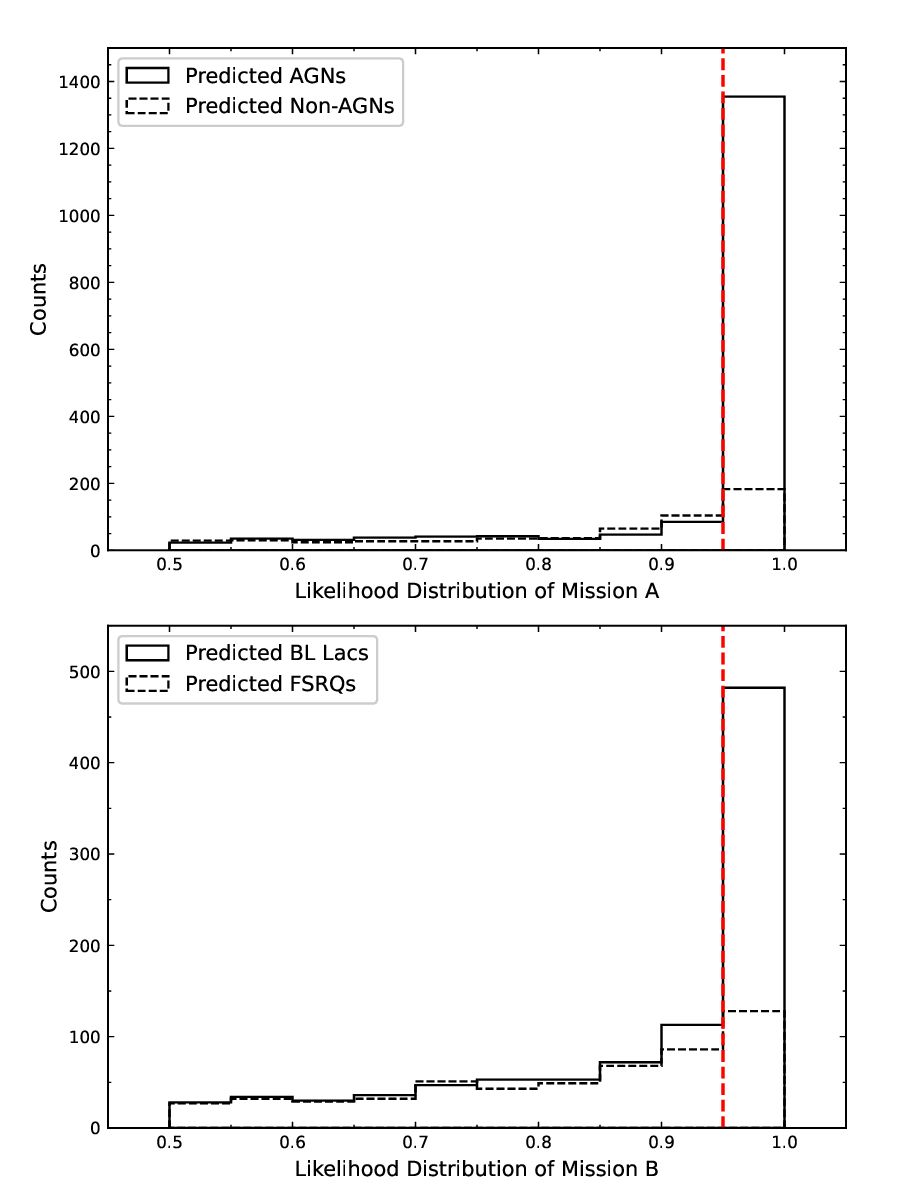}
\caption{The likelihood probability distribution of Mission A (upper panel) and Mission B (lower panel).
The AGN and Non-AGN candidates are shown in the upper histogram with the solid line and the dashed line, respectively.
The BL Lac and FSRQ candidates are shown in the lower histogram with the solid line and the dashed line, respectively.
The dashed red lines are the set boundary of likelihood probability equals 95\%.}
\label{dis_liki}
\end{figure}

By employing the proposed method, we {managed} to predict 2291 uncertain sources into {1731} AGNs and {560} Non-AGNs, and predict 1493 BCUs into {948} BL Lacs and {545} FSRQs.
The likelihood probabilities of these predicted uncertain sources and BCUs are shown in Tab. \ref{tab_prediction} and the distribution is displayed in Fig. \ref{dis_liki}.

We set a boundary of likelihood probability greater than 95\%, which is shown as dashed red lines in Fig. \ref{dis_liki}, to claim this source as a candidate of the corresponding class.
In this case, we get further constrained results into {1354} AGN candidates in Mission A, {482} BL Lacs candidates and {128} FSRQ candidates in Mission B, as shown in the last column of Tab. \ref{tab_prediction}.

There have been several methods employed to classify 4FGL BCUs as FSRQ candidates or BL Lac candidates, as Mission B of the present work, in the previous works.
\cite{Kang2019ApJ887} utilized three supervised ML methods (RF, SVM and ANN) to make a classification for 1312 4FGL\_DR1 BCUs, and carried out a combined classification result of 724 BL Lac candidates and 332 FSRQ candidates.
Crossing matching the results from \cite{Kang2019ApJ887} and our results, we found that there are {419} overlapping BCUs in two works.
Among the overlapping {419} BCUs, there are {324} BCUs predicted as BL Lac candidates and {90} BCUs are predicted as FSRQ candidates in both works, which gives our results a consistency of {98.8\%} with Kang's work.

Similarly, five different supervised ML algorithms (RF, LR, XGBoost, CatBoost, and neural network) were applied to the 4LAC\_DR3 BCUs in \cite{Agarwal2023ApJ946}, and they managed to classify 610 BCUs as BL Lac candidates and 333 BCUs as FSRQ candidates.
A comparison between our work and Agarwal's work suggests that there are {481} overlapping BCUs in the two works, {392} BCUs are classified as BL Lac candidates and {87} BCUs are classified as FSRQ candidates in both works, which gives our results a consistency of {99.6\%} with their work.

Besides, \cite{Fan2022Universe} employed three physical parameters and built diagrams among them (the photon spectrum index against the photon flux diagram; the photon spectrum index against the variability index diagram; the variability index against the photon flux) to separate the known BL Lacs from known FSRQs.
Then they used the boundary to divide BCUs into BL Lacs and FSRQs. 
In their work, 751 BCUs were classified as BL Lac candidates and 210 BCUs were classified as FSRQ candidates.
There are {492} overlapping BCUs in the two works, {409} BCUs are classified as BL Lac candidates and {83} BCUs are classified as FSRQ candidates in both works, which gives our results a consistency of 100\% with their work.

\section{Conclusion}
\label{sect_conclusion}
In this paper, the correlation features of attribute space of the 4FGL\_DR3 dataset are highlighted by the proposed FDIDWT method, and the intrinsic features hidden in data are further extracted by a lightweight MatchboxConv1D model. 
With the combination of the FDIDWT method and the MatchboxConv1D model, we have obtained the results with an accuracy of {$96.65\% \pm 1.32\%$} for Mission A and an accuracy of {$92.03\% \pm 2.2\%$} for Mission B.
As for the likelihood probability boundary 95\%, we {managed} to classify {1354} AGN candidates in Mission A, {482} BL Lacs candidates and {128} FSRQ candidates in Mission B.
A high consistency of greater than {98\%} emerges by comparing our predicted candidates with those from previous works.
More importantly, our method has the advantage of finding less variable and faint sources.

\section{Acknowledgments}
\label{sect_acknowledgments}
H.B.X acknowledges the support from the National Natural Science Foundation of China (NSFC) under grant No.12203034, from the Shanghai Science and Technology Fund under grant No. 22YF1431500, and from the science research grants from the China Manned Space Project.
Z.J.L acknowledges the support from NSFC grant 12141302, the Shanghai Science and Technology Fund under grant No. 20070502400, and from the science research grants from the China Manned Space Project.
J.H.F acknowledges the support of the NSFC U2031201, NSFC 11733001, the Scientific and Technological Cooperation Projects (2020–2023) between the People’s Republic of China and the Republic of Bulgaria, the science research grants from the China Manned Space Project with No. CMS-CSST-2021-A06, and the support for Astrophysics Key Subjects of Guangdong Province and Guangzhou City.

\bibliography{lib}{}
\bibliographystyle{aasjournal}

\end{document}